\documentclass[conference]{IEEEtran}
\IEEEoverridecommandlockouts
\usepackage[T1]{fontenc}
\usepackage[utf8]{inputenc}
\usepackage{textcomp}
\usepackage{graphicx}
\usepackage{booktabs}
\usepackage{array}
\usepackage{caption}
\usepackage{subfig}

\usepackage{url}
\usepackage[hidelinks]{hyperref}
\usepackage{balance}
\usepackage{xcolor}
\usepackage{needspace}
\usepackage{listings}
\begin{document}

\title{An Empirical Evaluation of Signing and Verification Performance\\ of Verifiable Credential Formats:\\ A Cross-Language Benchmark of SD-JWT VC, W3C VCDM, and mdoc}

\author{\IEEEauthorblockN{Naohiro Fujie}
\IEEEauthorblockA{\textit{MIRAI Design Laboratory}\\
\textit{ITOCHU Techno-Solutions Corporation}\\
\textit{OpenID Foundation Japan}\\
Tokyo, Japan\\
naohiro.fujie@ctc-g.co.jp}
\and
\IEEEauthorblockN{Shigeya Suzuki}
\IEEEauthorblockA{\textit{Global Research Institute}\\
\textit{Keio University}\\
Tokyo, Japan\\
shigeya@wide.ad.jp}}

\maketitle

\thispagestyle{plain}
\pagestyle{plain}

\begin{abstract}
We compare the signing and verification performance of three Verifiable Credential (VC) formats---SD-JWT VC; the W3C Verifiable Credentials Data Model 2.0 with a Data Integrity cryptosuite (W3C VCDM); and mdoc (ISO/IEC 18013-5)---over one scope: issue, present with selective disclosure, verify. Node.js, Go and Python implementations ran in four environments, one a cloud VM (N=2,000 across five runs; medians of each statistic and the spread). For an end-to-end exchange disclosing 5 of 20 attributes, the total (issue, present, verify) on Node.js in the cloud VM is 0.265\,ms for mdoc, 0.298\,ms for SD-JWT VC and 0.534\,ms for W3C VCDM. The four span a factor of five in speed; across them W3C VCDM takes 1.73--2.04$\times$ mdoc and 1.57--1.79$\times$ SD-JWT VC---at least five times our 10\% band---and the ordering survives unifying the algorithm on Ed25519 and P-256. The mdoc--SD-JWT VC difference is not robust: the totals differ by 12\%, 28\%, 19\% and 10\% in the four environments, and at verification mdoc is the slower in one of them. RDFC-1.0 canonicalization produces the gap: the cryptosuite canonicalizes twice per operation, so one exchange canonicalizes six times. We attribute that cost to the specification rather than to a library mainly by replacing RDFC-1.0 with JCS, which brings W3C VCDM verification to within 5\% of SD-JWT VC in three environments and within 12\% in the fourth; three implementations in three languages agree, though they share a lineage. Because eddsa-rdfc-2022 has no derivation step, presenting a subset re-signs it; ecdsa-sd-2023, which preserves the issuer signature, costs 3.4\,ms to issue and 1.5--2.4\,ms to derive and verify there. With production education schemas, canonicalization costs 31--49$\times$ a simple credential across the four, and blank node labeling is not the driver. The measured path ends at the issuer signature check.
\end{abstract}

\begin{IEEEkeywords}
Verifiable Credentials, SD-JWT, JSON-LD, mdoc, signature verification performance, RDF canonicalization, RDFC-1.0, benchmarking
\end{IEEEkeywords}

\section{Introduction}

This paper addresses three questions, from the standpoint of a deployment choosing a credential format for the verifier it will operate:

\begin{itemize}
\item \textbf{RQ1.} Over an identical scope---issue, present with selective disclosure, verify---how do the three formats compare, and does the ordering depend on the signature algorithm or on the machine?
\item \textbf{RQ2.} Where does any difference come from: the specification's required processing, the library implementing it, or the signature primitive?
\item \textbf{RQ3.} What does the comparison look like when the guarantee is held fixed---that is, when the W3C VCDM side must also preserve the issuer's signature across selective disclosure?
\end{itemize}

We answer these by measurement rather than analysis, and we are explicit about the boundary: the measured path runs from accepting an input to having checked the issuer signature over it. Key resolution (for mdoc, x5chain path validation), status checking, holder binding and transport are outside it, and Section~\ref{sec:disc-perf} states what that exclusion means for how the numbers should be used.

In the field of digital identity, Verifiable Credentials (VCs) have attracted attention as a mechanism that lets a verifier mechanically verify issuer-signed claims without querying a third party. The principal options available to implementers today are SD-JWT VC \cite{ref1}, configurations based on the W3C Verifiable Credentials Data Model 2.0 (VCDM 2.0) \cite{ref2}, and mdoc \cite{ref3}. These are not specifications at the same level of abstraction. SD-JWT VC and mdoc are credential formats that jointly specify claim structure, encoding and securing. VCDM 2.0, by contrast, specifies only an abstract data model for credentials; securing is supplied by separately combining it with a securing mechanism (Data Integrity \cite{ref4} or VC-JOSE-COSE \cite{ref5}). This paper evaluates the configuration that combines VCDM 2.0 with the Data Integrity cryptosuite eddsa-rdfc-2022; we refer to it as ``VCDM 2.0 + Data Integrity'' and abbreviate it as W3C VCDM. Where we mean the data model specification alone, without a securing mechanism, we write ``VCDM 2.0'', to distinguish it from the configuration evaluated here. The three rest on different design philosophies, so selecting among them requires multifaceted evaluation.

SD-JWT VC is specified as an IETF Internet-Draft under active development and adopts JWT compact serialization. It has strong affinity with the existing OAuth 2.0/OpenID Connect ecosystem, and selective disclosure is provided by the SD-JWT specification \cite{ref6}. mdoc is the format for mobile identification documents specified in ISO/IEC 18013-5 \cite{ref3} and adopts binary encoding based on CBOR (Concise Binary Object Representation) and COSE (CBOR Object Signing and Encryption). VCDM 2.0 + Data Integrity, by contrast, expresses the credential data model in JSON-LD, and the Data Integrity \cite{ref4} cryptosuite eddsa-rdfc-2022 requires RDF dataset canonicalization (RDFC-1.0, the W3C Recommendation of the algorithm formerly developed as URDNA2015) \cite{ref7} for signing and verification.

Empirical studies that compare these three formats under identical conditions remain limited. Moreover, EU Regulation 2024/1183 (eIDAS 2.0) \cite{ref8} establishes the European Digital Identity Wallet (EUDIW), and the implementing acts adopted under it designate SD-JWT VC and mdoc as the credential formats wallets must support, increasing the practical importance of quantitative comparison between formats. With a benchmark tool that we developed for this study and released as open source, this paper performs a reproducible comparative evaluation along two axes. The first is a signing/verification speed benchmark that measures the time per operation of signing and verification, including attribute-count scaling, selective disclosure, and canonicalization cost for production schemas. The second is an implementation complexity analysis that measures lines of code, cyclomatic complexity, the number of asynchronous steps, and the number of external network calls. Security-relevant attack surfaces that arise from the verification pipeline---denial of service (DoS), server-side request forgery (SSRF), context injection, and so on---are analyzed in a companion paper by the authors (arXiv:2611.04461) that uses the same measurement infrastructure.

Note that the performance comparison in this paper adopts the signature algorithms standardly used in each format's ecosystem (Ed25519 for SD-JWT VC and W3C VCDM, ECDSA P-256 for mdoc), so the measurements reflect not only differences in verification pipelines (canonicalization and preprocessing) but also differences in signature algorithms. This confounding factor is discussed in detail in Section~\ref{sec:disc} and in Section~\ref{sec:ttv}.

\section{Related Work}

Several prior studies compare VC formats, but work that quantitatively evaluates signature verification performance in a single environment, and that confirms reproducibility across multiple language runtimes, remains limited. Table~\ref{tab:related} positions this paper against the closest work along the axes that matter for the comparison: whether all three formats are measured by one method, whether selective disclosure is compared at equal guarantee, how many language runtimes are covered, and whether canonicalization is separated from the rest of the pipeline. We searched IEEE Xplore, the ACM Digital Library, SpringerLink, ScienceDirect and arXiv in October 2026 for combinations of \emph{verifiable credential}, \emph{SD-JWT}, \emph{mdoc}, \emph{mDL}, \emph{ISO 18013-5}, \emph{Data Integrity} and \emph{benchmark}/\emph{performance}/\emph{evaluation}, and followed citations from the records retrieved. Within that scope we found no study reporting measured signing or verification time for mdoc; we state the scope because the claim is otherwise unfalsifiable, and we do not claim exhaustiveness.

\begin{table*}[!t]
\centering
\caption{Scope of the closest related work. ``All three by one method'' means SD-JWT VC, W3C VCDM and mdoc measured in one environment under one methodology. ``Equal-guarantee disclosure'' means that the selective disclosure mechanisms compared preserve the issuer signature alike.}\label{tab:related}
\footnotesize\setlength{\tabcolsep}{4pt}
\begin{tabular}{@{}>{\raggedright\arraybackslash}p{0.17\textwidth} >{\raggedright\arraybackslash}p{0.15\textwidth} >{\raggedright\arraybackslash}p{0.15\textwidth} >{\raggedright\arraybackslash}p{0.13\textwidth} >{\raggedright\arraybackslash}p{0.17\textwidth} >{\raggedright\arraybackslash}p{0.14\textwidth}@{}}
\toprule
\textbf{Study} & \textbf{All three by one method} & \textbf{Equal-guarantee disclosure} & \textbf{Runtimes / machines} & \textbf{Canonicalization separated} & \textbf{Implementation complexity}\tabularnewline
\midrule
Brunner et al. \cite{ref9} & no (no measurement) & --- & --- & no & no\tabularnewline
Wu and Tian \cite{ref10} & no (BBS+ vs SD-JWT) & yes, within that pair & --- (literature review) & no & no\tabularnewline
Buldini et al. \cite{ref11} & no (SD-JWT vs CSD-JWT) & yes, within that pair & 1 / 1 & no & no\tabularnewline
Lagos et al. \cite{ref12} & no (SD-JWT vs.\ BBS) & yes, within that pair & several / 4 devices & no & no\tabularnewline
\textbf{This paper} & \textbf{yes} & \textbf{yes} (Table~\ref{tab:sd}) & \textbf{3 / 4} & \textbf{yes} (Tables~\ref{tab:breakdown}, \ref{tab:prim}) & \textbf{reference implementations only}\tabularnewline
\bottomrule
\end{tabular}
\end{table*}

This section surveys the major related work and clarifies the positioning summarized above.

Brunner et al. \cite{ref9} comprehensively surveyed the technical landscape of Decentralized Identifiers (DIDs) and VCs and organized the workflows of decentralized identification and verification. Their study, however, includes no quantitative performance comparison across formats and remains at the conceptual level.

Regarding the comparison of selective disclosure mechanisms, Wu and Tian \cite{ref10} conducted a systematic literature review (SLR) of BBS+ signatures and SD-JWT, extracting and analyzing 31 primary studies published between 2017 and 2025 from IEEE, ACM, SpringerLink, and other repositories. They report that BBS+ derived proofs have a fixed size of roughly 140 bytes and verify in about 12\,ms, whereas SD-JWT presentations grow with the number of disclosed claims but still verify in under 10\,ms when two claims are disclosed. Their focus, however, is the comparison of selective disclosure primitives (BBS+ vs. SD-JWT), which is a different evaluation axis from the baseline signing and verification performance of the three formats studied here (SD-JWT VC, W3C VCDM, and mdoc).

Buldini et al. \cite{ref11} proposed a compact selective disclosure scheme (CSD-JWT) based on cryptographic accumulators, showing up to 46\% lower memory usage and up to 93\% smaller Verifiable Presentations compared with SD-JWT. Their work targets resource-constrained hardware wallets and reports that issuance completes within a few milliseconds even for credentials with 100 claims. Their evaluation, however, is limited to a comparison of SD-JWT and CSD-JWT and does not cover W3C VCDM or mdoc.

For RDF dataset canonicalization, the RDFC-1.0 specification \cite[\S7.1]{ref7} documents the worst-case complexity of RDFC-1.0: for graphs containing blank nodes, the cost of canonical labeling can grow superlinearly with input size. On the theoretical side, Hogan \cite{ref13} shows that labeling blank nodes requires exponential time in the worst case, while a survey of 9.9 million real-world graphs finds such hard cases to be rare. This paper quantifies the extent to which this specification-level characteristic manifests as implementation-level verification time, using attribute-count scaling and production schemas.

For performance evaluation targeting the EUDIW, Lagos et al. \cite{ref12} present an extensible benchmark that measures SD-JWT and BBS-based selective disclosure across four devices---a desktop, a smartphone, a smartwatch and a single-board computer---and across implementations in different languages, motivated by the three-way unlinkability requirement of the EU regulation. Their axis is therefore the disclosure primitive rather than the credential format, and they report that the ranking depends on hardware, scheme and implementation language alike. This paper differs in both target and viewpoint: rather than comparing selective disclosure primitives, we measure the verification pipelines of three credential formats---SD-JWT VC, VCDM 2.0 + Data Integrity, and mdoc---across three language runtimes under an identical methodology, and we accompany the measurements with a decomposition of canonicalization cost and with implementation complexity metrics. To our knowledge no academic study quantifies the signing and verification performance of mdoc, and this paper also fills that gap.

Against this background, the contributions of this paper are threefold. First, we measure SD-JWT VC, W3C VCDM, and mdoc under an identical methodology and with the same measurement code across Node.js, Go, and Python in four execution environments spanning a cloud VM, two bare-metal x86\_64 machines of different generations and an arm64 desktop machine, and we show empirically which of the differences among formats survive a change of runtime and of machine and which do not---a distinction a single environment cannot draw. Second, in addition to a simple credential, we target attribute-count scaling and production education schemas, quantifying the conditions under which RDFC-1.0 canonicalization becomes the dominant cost of the verification pipeline. Third, we release as open source a reproducible measurement framework in which the statistical treatment (outlier handling, representative statistic, repeated independent runs) is unified across all three languages. This paper does not aim to render a definitive superiority judgment among VC formats; its main purpose is to provide quantitative data that implementers and standardization stakeholders can consult when selecting a format.

\section{Technical Specifications of Each Format}

\subsection{Overview}

Table~\ref{tab:t1} shows the basic configuration of the three formats evaluated in this paper. Each format differs in its data representation, securing mechanism, signature algorithm and canonicalization requirement, and these differences directly affect each evaluation axis of performance and complexity. The following sections detail the technical specifications, data models, and signing/verification processing methods and procedures for each format.

\begin{table*}[!t]
\centering
\caption{Basic configuration of the evaluated formats.}\label{tab:t1}
\footnotesize\setlength{\tabcolsep}{4pt}
\begin{tabular}{@{}l p{2.9cm} p{3.4cm} l p{2.3cm} p{3.6cm}@{}}
\toprule
\textbf{Format} & \textbf{Data Representation} & \textbf{Securing Mechanism} & \textbf{Signature Algorithm} & \textbf{Canonicalization} & \textbf{Primary Standard}\tabularnewline
\midrule
SD-JWT VC & JSON claims (JWS Compact + disclosure list) & Enveloping (JWS: JSON Web Signature) & EdDSA (Ed25519) & None & draft-ietf-oauth-sd-jwt-vc-22 \cite{ref1} (selective disclosure: \cite{ref6})\tabularnewline
W3C VCDM & JSON-LD (RDF data model) & Embedded proof (Data Integrity, eddsa-rdfc-2022) \cite{ref4} & Ed25519 & RDFC-1.0 & W3C VCDM 2.0 \cite{ref2}\tabularnewline
mdoc & CBOR & COSE\_Sign1 (MSO: Mobile Security Object) & ECDSA P-256 (ES256) & None & ISO/IEC 18013-5 \cite{ref3}\tabularnewline
\bottomrule
\end{tabular}

\vspace{2pt}
{\footnotesize The digest algorithm is SHA-256 in all three formats: \texttt{\_sd\_alg} in SD-JWT VC, the Hashing Algorithm of Data Integrity, and the MSO digest algorithm in mdoc.}
\end{table*}

\subsection{SD-JWT VC}

\subsubsection{Standards Position}

SD-JWT VC (SD-JWT-based Verifiable Credentials) is a credential format specified as an IETF Internet-Draft, draft-ietf-oauth-sd-jwt-vc-22 \cite{ref1}, and still under active development. It adopts JSON Web Token (JWT, RFC 7519 \cite{ref14}) and JSON Web Signature (JWS, RFC 7515 \cite{ref15}) as its foundational technologies, and uses SD-JWT (Selective Disclosure for JWTs, RFC 9901 \cite{ref6}) for selective disclosure. It is characterized by high affinity with the existing OAuth 2.0/OpenID Connect ecosystem.

\subsubsection{Data Model}

SD-JWT VC data is represented in JWT Compact Serialization format. A JWT is a string formed by base64url-encoding three parts---Header, Payload, and Signature---and concatenating them with dots (.). The Header contains the algorithm identifier (alg) and token type (typ), while the Payload stores the credential claims.

The key claims defined in SD-JWT VC are as follows: iss (Issuer) is the issuer identifier, sub (Subject) is the credential subject identifier, and vct (Verifiable Credential Type) indicates the credential type. iat (Issued At) and exp (Expiration Time) are timestamps indicating the issuance time and expiration time. In addition, credential-specific claims (name, date of birth, etc.) may be included.

\Needspace*{5\baselineskip}
\textbf{Listing 1. SD-JWT VC Data Structure Example}\nopagebreak

\begin{lstlisting}
// Header (Base64url encoded)
{ "alg": "EdDSA", "typ": "vc+sd-jwt" }
// Payload (Base64url encoded)
{
  "iss": "https://issuer.example.com",
  "sub": "did:example:holder123",
  "vct": "https://credentials.example.com/identity",
  "iat": 1714000000,
  "exp": 1714003600,
  "given_name": "Taro",
  "family_name": "Yamada",
  "birthdate": "1990-01-01"
}
// Compact Serialization: Header.Payload.Signature
eyJhbGciOiJFZERTQSIsInR5cCI6InZjK3NkLWp3dCJ9.eyJpc3MiOi...._dBjftJeZ4CVP-mB92K27uhbUJU1p1r_wW1gFWFOEjXk
\end{lstlisting}

The \texttt{typ} value in Listing~1 is the one our measurement used. The draft has since replaced it: in draft-22 \cite{ref1} the media type is \texttt{application/dc+sd-jwt} and the header \texttt{typ} must be \texttt{dc+sd-jwt}, \texttt{dc} standing for ``digital credential''. We report the value as run rather than as currently specified. The substitution cannot affect any figure in this paper: both headers serialize to 33 bytes and Base64url-encode to the same 44 characters, so the signing input is identical in length, and the \texttt{typ} value is not otherwise read on the measured path.

\subsubsection{Signing Process}

The signing process for SD-JWT VC follows the JWS specification as follows. First, the Header is constructed and Base64url-encoded. Next, the Payload is constructed and Base64url-encoded. The Signing Input is formed as \texttt{Base64url(Header) || '.' || Base64url(Payload)}, and a signature is generated over this input using the private key. Finally, the signature value is Base64url-encoded, and the compact serialization \texttt{Header.Payload.Signature} is output.

The available signature algorithms come from the JOSE registry (JWA, RFC 7518 \cite{ref16}; EdDSA added by RFC 8037 \cite{ref17}) and include EdDSA (Ed25519/Ed448), ES256 (ECDSA using P-256 and SHA-256), ES384 and ES512. This experiment used EdDSA (Ed25519) \cite{ref18}. The fact that the entire signing process completes with a single API call (e.g., SignJWT.sign() in jose) directly contributes to implementation simplicity compared to other formats.

\subsubsection{Verification Process}

The verification process also follows the JWS specification. The token is split into three parts by dots, and the Header is decoded to obtain the alg header parameter. The verifier checks the alg against an allowlist of acceptable algorithms and rejects algorithms not on the list. This is a defense measure recommended in RFC 8725 \cite{ref19} (JWT Best Current Practices) and serves as a countermeasure against alg:none attacks and algorithm confusion attacks. Next, the signing input is reconstructed, and the signature is verified using the issuer's public key. If the signature is valid, the Payload is decoded and the claims are checked. Of the registered claims the specification lists, only \texttt{vct} is REQUIRED; \texttt{iss}, \texttt{exp}, \texttt{nbf} and \texttt{status} are OPTIONAL, and \texttt{cnf} is REQUIRED only when cryptographic Key Binding is supported. Our reference verifier checks \texttt{vct}, and also \texttt{iss} because the measured configuration conveys the issuer that way rather than through an \texttt{x5c} header.

\subsubsection{Selective Disclosure}

Selective disclosure in SD-JWT VC is provided by the SD-JWT specification \cite{ref6}. The issuer replaces disclosable claims in the Payload with SHA-256 hashes and provides the original values separately as Disclosures. A Disclosure is built by creating a JSON array whose three elements are, in this order, a salt, the claim name and the claim value, and then base64url-encoding the UTF-8 byte sequence of that array \cite{ref6}. It is the JSON representation itself that is encoded, not a concatenation of the three values. Because the digest is computed over the base64url string itself, no canonicalization is required to absorb differences in whitespace or Unicode escaping. For the claim given\_name, for example, the Disclosure is \texttt{base64url(["<salt>","given\_name","Taro"])} and only its SHA-256 digest is stored in the \_sd array of the Payload. The holder selects and sends only the Disclosures to be revealed to the verifier. The verifier confirms the authenticity of disclosed claims by matching the hash of each Disclosure against the corresponding hash in the Payload. This mechanism achieves selective disclosure without canonicalization processing.

\subsection{W3C VCDM 2.0 + Data Integrity}

\subsubsection{Standards Position}

The W3C Verifiable Credentials Data Model 2.0 (VCDM 2.0) \cite{ref2} is a specification that expresses the credential data model in JSON-LD; it does not itself define a signature mechanism. It provides semantic interoperability through JSON-LD (JSON for Linked Data) and provides semantic typing based on the RDF (Resource Description Framework) graph model. Cryptographic suites defined in the W3C Data Integrity specification \cite{ref4} are used for signing. This experiment adopted the eddsa-rdfc-2022 cryptographic suite (transformation: RDFC-1.0, hash: SHA-256, signature: EdDSA Ed25519). Ed25519Signature2020 is a legacy Linked Data Proofs cryptographic suite, while eddsa-rdfc-2022 is its successor defined in the Data Integrity EdDSA Cryptosuites specification \cite{ref20}. Both share RDFC-1.0 canonicalization and Ed25519 signing, so their canonicalization and signing pipeline performance characteristics are equivalent.

\subsubsection{Data Model}

The data structure of W3C VCDM consists of vocabulary definitions via @context, type, issuer, credentialSubject, and proof (signature information). @context is a core concept of JSON-LD, containing references to external vocabulary definitions that map JSON keys (property names) to IRIs (Internationalized Resource Identifiers). This ensures that even VCs issued by different issuers have their field semantics unambiguously interpreted by machine, as long as they share the same @context.

\Needspace*{5\baselineskip}
\textbf{Listing 2. W3C VCDM Data Structure Example}\nopagebreak

\begin{lstlisting}
{
  "@context": [
    "https://www.w3.org/ns/credentials/v2"
  ],
  "id": "https://example.com/credentials/1872",
  "type": "VerifiableCredential",
  "issuer": "https://issuer.example.com",
  "validFrom": "2024-01-01T00:00:00Z",
  "credentialSubject": {
    "id": "did:example:holder123",
    "given_name": "Taro",
    "family_name": "Yamada",
    "birthdate": "1990-01-01"
  },
  "proof": {
    "type": "DataIntegrityProof",
    "cryptosuite": "eddsa-rdfc-2022",
    "created": "2024-01-01T00:00:00Z",
    "verificationMethod": "did:example:issuer#key-1",
    "proofPurpose": "assertionMethod",
    "proofValue": "z58DAdFfa9SkqZMVP..."
  }
}
\end{lstlisting}

Listing 2 shows a conforming VCDM 2.0 credential. The credential measured in this paper embeds a minimal inline \texttt{@context} that defines only the terms it uses, in place of the context URL, and carries the term \texttt{issuanceDate}; Section~\ref{sec:bench-speed} states what that excludes from the measured cost.

\subsubsection{Data Integrity Proof Specification}

The W3C Verifiable Credential Data Integrity 1.0 specification \cite{ref4} defines a general-purpose framework for generating and verifying cryptographic proofs for VCs. The specification externalizes the selection of specific cryptographic algorithms as cryptographic suites (cryptosuite), while the framework itself specifies a three-stage pipeline independent of cryptographic suites: Transformation, Hashing, and Proof Generation/Verification.

\textbf{The proof object.} In Data Integrity, signature information is embedded in the VC as a proof object. The key properties of the proof object are those of Table~\ref{tab:t2}.

\begin{table}[!t]
\centering
\caption{Key properties of the \texttt{proof} object (Data Integrity 1.0).}\label{tab:t2}
\footnotesize\setlength{\tabcolsep}{4pt}
\begin{tabular}{@{}l@{\hspace{4pt}}l@{\hspace{4pt}}>{\raggedright\arraybackslash}p{0.44\columnwidth}@{}}
\toprule
\textbf{Property} & \textbf{Required} & \textbf{Description}\tabularnewline
\midrule
type & Required & Proof type (e.g., DataIntegrityProof)\tabularnewline
cryptosuite & Required & Cryptographic suite identifier (e.g., eddsa-rdfc-2022)\tabularnewline
verificationMethod & Optional & Identifier of the public key (verification method) used for verification, expressed as a DID URL or an HTTPS URL. Optional in the specification: when absent, the key is obtained by other means, for instance through \texttt{cryptosuite}. Present in every configuration measured here\tabularnewline
proofPurpose & Required & Proof purpose (assertionMethod / authentication, etc.)\tabularnewline
created & Optional & Proof creation datetime (xsd:dateTime)\tabularnewline
domain & Optional & Valid domain for the proof (replay attack prevention)\tabularnewline
challenge & Optional & Challenge value (replay attack prevention)\tabularnewline
proofValue & Required & Signature value (Multibase + Base58btc encoding)\tabularnewline
\bottomrule
\end{tabular}
\end{table}

\subsubsection{Cryptographic Suite Structure}

The Data Integrity specification requires a cryptographic suite specification to detail four algorithms. First, the Transformation Algorithm converts the input document to a canonical form suitable for hashing. For eddsa-rdfc-2022, this is RDFC-1.0 canonicalization, which converts the JSON-LD document to an RDF dataset and generates canonical N-Quads through blank node relabeling. Second, the Hashing Algorithm computes a hash of the canonicalized data. SHA-256 is used. Third, the Proof Serialization Algorithm generates the final proof value: Ed25519 signing is performed and the result is Multibase-encoded. Fourth, the Proof Verification Algorithm checks that value against the hash and the verification method.

The cryptographic suites the W3C Verifiable Credentials Working Group has published are listed in Table~\ref{tab:t3}. All but \texttt{bbs-2023} are Recommendations; \texttt{bbs-2023} is a Candidate Recommendation Draft at the time of writing.

\begin{table*}[!t]
\centering
\caption{W3C Data Integrity cryptographic suites.}\label{tab:t3}
\footnotesize\setlength{\tabcolsep}{4pt}
\begin{tabular}{@{}lllll@{}}
\toprule
\textbf{Cryptographic Suite} & \textbf{Transformation} & \textbf{Hash} & \textbf{Sign} & \textbf{Specification}\tabularnewline
\midrule
eddsa-rdfc-2022 & RDFC-1.0 & SHA-256 & EdDSA (Ed25519) & VC DI EdDSA 1.0 \cite{ref20}\tabularnewline
eddsa-jcs-2022 & JCS (JSON Canonicalization Scheme, RFC 8785) \cite{ref21} & SHA-256 & EdDSA (Ed25519) & VC DI EdDSA 1.0 \cite{ref20}\tabularnewline
ecdsa-rdfc-2019 & RDFC-1.0 & SHA-256/384 & ECDSA (P-256/P-384) & VC DI ECDSA 1.0 \cite{ref22}\tabularnewline
ecdsa-jcs-2019 & JCS (RFC 8785) & SHA-256/384 & ECDSA (P-256/P-384) & VC DI ECDSA 1.0 \cite{ref22}\tabularnewline
ecdsa-sd-2023 & RDFC-1.0 + selective disclosure & SHA-256 & ECDSA (P-256) & VC DI ECDSA 1.0 \cite{ref22}\tabularnewline
bbs-2023 & RDFC-1.0 + BBS-specific & SHA-256 & BBS (BLS12-381) & VC DI BBS 1.0\tabularnewline
\bottomrule
\end{tabular}
\end{table*}

The eddsa-rdfc-2022 cryptographic suite used in this experiment adopts RDFC-1.0 canonicalization for transformation, SHA-256 for hashing, and EdDSA (Ed25519) for signing. In contrast, eddsa-jcs-2022 adopts JSON Canonicalization Scheme (JCS, RFC 8785) \cite{ref21} instead of RDFC-1.0, avoiding its computational complexity issues since it does not perform RDF graph canonicalization. However, JCS-based suites do not provide RDF-based semantic interoperability, departing from the original design intent of W3C VCDM. Additionally, bbs-2023---still a Candidate Recommendation Draft---is a suite that provides zero-knowledge proof-based selective disclosure and presentation unlinkability using the BBS signature scheme.

\subsubsection{Signing Process (Add Proof in Data Integrity; Proof Serialization in the cryptosuite)}

The signing process based on the Data Integrity specification's Add Proof algorithm consists of the following steps. We explain using the eddsa-rdfc-2022 cryptographic suite as an example.

\textbf{Step 1: Constructing proof options.} Construct a proof options object from metadata such as type, cryptosuite, verificationMethod, proofPurpose, and created. This object does not include the proofValue field (since signing has not yet occurred).

\textbf{Step 2: Document transformation.} Apply the RDFC-1.0 canonicalization algorithm to the input document with the proof property removed. First, resolve the JSON-LD context based on @context and expand the document into an RDF dataset. Then, assign canonical labels to blank nodes in the RDF dataset and output a canonicalized N-Quads format string. This canonicalization process can be the most computationally expensive step, as it determines unique labels for all blank nodes based on graph structure.

\textbf{Step 3: Canonicalization and hashing of proof options.} The proof options object constructed in Step 1 is also treated as a JSON-LD document and similarly undergoes RDFC-1.0 canonicalization. A SHA-256 hash is computed over the canonicalized proof options string to obtain the 32-byte proofOptionsHash (proof configuration hash). This step ensures that the proof options content is also protected by the signature.

\textbf{Step 4: Document hashing.} Compute a SHA-256 hash over the canonicalized document string obtained in Step 2 to obtain the 32-byte transformedDocumentHash.

\textbf{Step 5: Hash data concatenation.} Concatenate proofOptionsHash (32 bytes) and transformedDocumentHash (32 bytes) to generate 64-byte hashData. This concatenation ensures that both the proof options and the document body are protected by a single signature.

\textbf{Step 6: Signature generation.} Generate an EdDSA signature over hashData (64 bytes) using the Ed25519 private key. The signature value is stored in the proofValue field as Multibase encoding (Base58btc header \texttt{z} + Base58btc encoding).

\textbf{Step 7: Appending the proof.} Append the proof object containing proofValue to the credential and output the signed VC.

\Needspace*{5\baselineskip}
\textbf{Listing 3. Conceptual Flow of Data Integrity Signing Process}\nopagebreak

\begin{lstlisting}
// Add Proof Algorithm (eddsa-rdfc-2022)
// 1. Construct proof options (excluding proofValue)
proofOptions = {
  type: "DataIntegrityProof",
  cryptosuite: "eddsa-rdfc-2022",
  verificationMethod: "did:example:issuer#key-1",
  proofPurpose: "assertionMethod",
  created: "2024-01-01T00:00:00Z"
}
// 2. Document Transformation
// RDFC-1.0 canonicalization -> N-Quads string
canonicalDoc = RDFC_1_0.canonicalize(
  document_without_proof)
// 3. Canonicalize proof options + hash
canonicalOpts = RDFC_1_0.canonicalize(proofOptions)
proofOptionsHash = SHA256(canonicalOpts) // 32 bytes
// 4. Hash the document
docHash = SHA256(canonicalDoc) // 32 bytes
// 5. Concatenate hash data
hashData = proofOptionsHash || docHash // 64 bytes
// 6. Generate EdDSA signature
signature = Ed25519.sign(hashData, privateKey)
// 7. Multibase-encode as proofValue
proofOptions.proofValue = "z" +
Base58btc.encode(signature)
\end{lstlisting}

An important characteristic of this processing flow is that canonicalization is executed twice: once for the document body and once for the proof options. This ensures that during verification, not only the document content but also the proof metadata (creation date, verification method, purpose, etc.) is covered by the signature. The Sig\_Structure in COSE\_Sign1, which includes the protected header in the signing target, serves a similar role, but W3C VCDM requires explicit canonicalization of both the document and proof options, adding processing overhead.

\subsubsection{Verification Process (Verify Proof in Data Integrity; Proof Verification in the cryptosuite)}

The Data Integrity specification's Verify Proof algorithm is structured as the inverse operation of Add Proof. The verifier executes the following steps.

\textbf{Step 1: Separating the proof.} Extract the proof object from the VC, obtain and retain the proofValue (signature value), then remove proofValue from the proof object to reconstruct proof options.

\textbf{Step 2: Document transformation and hashing.} Following the same procedure as signing, perform RDFC-1.0 canonicalization on the document excluding proof, and compute the SHA-256 hash (transformedDocumentHash).

\textbf{Step 3: Canonicalization and hashing of proof options.} Perform RDFC-1.0 canonicalization on proof options and compute the SHA-256 hash (proofOptionsHash).

\textbf{Step 4: Hash data concatenation and signature verification.} Reconstruct hashData (64 bytes) by concatenating proofOptionsHash and transformedDocumentHash. Obtain the signature value by Multibase-decoding proofValue, and verify the EdDSA signature using the public key identified by verificationMethod.

An important characteristic of the verification process is the network dependency of canonicalization. HTTP requests are generated to URLs specified in @context, so network latency and context server availability directly affect verification processing time. This characteristic also introduces security risks such as SSRF and DoS, which the companion paper analyzes. Using a static context loader (local cache) can eliminate network dependency, but requires explicit configuration by application developers.

\subsubsection{RDFC-1.0 Canonicalization Algorithm Details}

In the Data Integrity signing and verification pipeline, RDFC-1.0 \cite{ref7} used in the transformation stage is an algorithm for canonicalization of RDF datasets. The purpose of canonicalization is to guarantee that two semantically equivalent RDF graphs have the same string representation. This ensures that for VCs with the same content, the same hash value is always obtained regardless of differences in representation (property order, whitespace, blank node label differences, etc.).

The RDFC-1.0 algorithm consists of three major phases. In the first phase (Initialization), all blank nodes in the input RDF dataset are enumerated, and an initial map computing hash values for each blank node is constructed. In the second phase (Hash First Degree Quads), for each blank node, hash values of the quads directly referenced by that blank node are computed. In the third phase (Hash N-Degree Quads), for blank nodes not identified in the second phase, recursive hash computation considering a wider range of quads is performed, assigning a unique canonical label to all blank nodes.

The recursive hash computation in the third phase is what causes exponential computational complexity in the worst case. When a cyclic structure (poison graph) where all blank nodes mutually reference each other is input, combinatorial explosion occurs in the N-Degree Quads hash computation. \S4.4.3 of the W3C RDFC-1.0 specification states that implementations MUST defend against potential denial-of-service attacks by raising suitable exceptions and terminating early, and a note there suggests placing a limit (call limit) on the number of Hash N-Degree Quads (\S4.8) invocations based on the number of blank nodes. In the implementations we measured, however, no such limit is active by default: in rdf-canonize@3.4.0, which jsonld@8.3.3 uses internally, the invocation limit \texttt{maxDeepIterations} defaults to \texttt{Infinity}, so no limit applies unless the caller sets one explicitly, and neither PyLD (3.3.0, the version measured here) nor json-gold provides any such mechanism. The call limit is therefore a parameter the verifier has to set, not one the library supplies. Considering the worst-case computational complexity characteristics noted in the RDFC-1.0 specification, call limits are important as a defensive implementation measure; their effectiveness is evaluated in the companion paper.

\subsubsection{Semantic Interoperability}

The greatest technical feature of W3C VCDM is RDF-based semantic interoperability. Since all properties are mapped to globally unique IRIs via @context, the meaning of fields never becomes ambiguous when exchanging credentials between different ecosystems. For example, the field \texttt{issuer} is resolved to the IRI \texttt{https://www.w3.org/2018/credentials\#issuer} through the W3C VC context and is interpreted as the same meaning in any system. This property plays an essential role in credential interoperability within the Decentralized Identity (DID) ecosystem. However, as a trade-off for this semantic interoperability, the aforementioned canonicalization processing incurs performance and security overhead.

\subsection{mdoc (ISO/IEC 18013-5)}

\subsubsection{Standards Position}

mdoc is a credential format for mobile identity documents specified in ISO/IEC 18013-5:2021 \cite{ref3}. It adopts binary encoding via CBOR \cite{ref23} and signing via COSE \cite{ref24}, and is designed with offline presentation via short-range communication (NFC, BLE, Wi-Fi Aware) as its primary use case. ISO/IEC TS 18013-7 \cite{ref25} specifies extensions for online presentation (OID4VP).

\subsubsection{Data Model}

ISO/IEC 18013-5 defines the mdoc data model in CDDL (Concise Data Definition Language, RFC 8610 \cite{ref26}). The structure relevant here is Document, composed of docType (document type), issuerSigned (issuer-signed data) and, on presentation, deviceSigned. issuerSigned contains nameSpaces (data elements per namespace) and issuerAuth (issuer authentication data).

Individual data elements are stored as IssuerSignedItem structures. An IssuerSignedItem consists of four fields: digestID (digest identifier), random (random value of at least 16 bytes), elementIdentifier (data element identifier), and elementValue (data element value). The purpose of the random field is to prevent inference of data element contents from digest values alone, contributing to privacy protection during selective disclosure.

The three formats differ markedly in how they convey the credential type and the meaning of claim names. In VCDM 2.0 the type property carries the type, while @context binds claim names to an external vocabulary document. In SD-JWT VC the vct claim carries the type, and the meaning of claim names is left to the type definition it denotes, so no external document is resolved at verification time. mdoc has no counterpart to either: the type is given by docType and the claim namespace structurally by the keys of nameSpaces. Only VCDM 2.0 + Data Integrity therefore resolves an external context and canonicalizes the expanded result during signing and verification, and that difference is the source of the processing-cost gap measured in this paper.

\Needspace*{5\baselineskip}
\textbf{Listing 4. mdoc Data Structure (CDDL Definition, from ISO/IEC 18013-5)}\nopagebreak

\begin{lstlisting}
Document = {
  "docType" : DocType,
  "issuerSigned" : IssuerSigned
}
IssuerSigned = {
  "nameSpaces" : IssuerNameSpaces,
  "issuerAuth" : IssuerAuth ; COSE_Sign1
}
IssuerNameSpaces = {
  + NameSpace => [ + IssuerSignedItemBytes ]
}
IssuerSignedItemBytes = #6.24(bstr .cbor IssuerSignedItem)
IssuerSignedItem = {
  "digestID" : uint,
  "random" : bstr, ; min 16 bytes
  "elementIdentifier" : DataElementIdentifier,
  "elementValue" : DataElementValue
}
\end{lstlisting}

\subsubsection{Mobile Security Object (MSO)}

mdoc signing is performed through the Mobile Security Object (MSO). The MSO is a structure that aggregates digest values of all data elements and is stored as the payload of IssuerAuth (COSE\_Sign1). The key fields of the MSO are as follows: version stores the MSO structure version (currently ``1.0''), digestAlgorithm stores the digest algorithm identifier (one of SHA-256/SHA-384/SHA-512), valueDigests stores the map of digest IDs to digest values per namespace, deviceKeyInfo stores the public key for mdoc authentication, docType stores the document type, and validityInfo stores the signing date and validity period information.

\Needspace*{5\baselineskip}
\textbf{Listing 5. MSO Structure Definition (from ISO/IEC 18013-5 \S9.1.2.4)}\nopagebreak

\begin{lstlisting}
IssuerAuth = COSE_Sign1
; payload = MobileSecurityObjectBytes
MobileSecurityObjectBytes =
#6.24(bstr .cbor MobileSecurityObject)
MobileSecurityObject = {
  "version" : tstr, ; "1.0"
  "digestAlgorithm" : tstr, ; "SHA-256" etc.
  "valueDigests" : ValueDigests,
  "deviceKeyInfo" : DeviceKeyInfo,
  "docType" : tstr,
  "validityInfo" : ValidityInfo
}
ValueDigests = {
  + NameSpace => DigestIDs
}
DigestIDs = {
  + DigestID => Digest ; DigestID = uint, Digest = bstr
}
ValidityInfo = {
  "signed" : tdate,
  "validFrom" : tdate,
  "validUntil" : tdate,
  ? "expectedUpdate" : tdate
}
\end{lstlisting}

\subsubsection{Signing Process}

The mdoc signing (issuance) process follows the procedure specified in ISO/IEC 18013-5 \S9.1.2.

\textbf{Stage 1: CBOR encoding of data elements and digest computation.} For each data element (elementIdentifier + elementValue), assign a random salt (random) of at least 16 bytes and a digestID to construct an IssuerSignedItem structure, then CBOR-encode it. For each encoded IssuerSignedItemBytes, individually compute a SHA-256 (or SHA-384/SHA-512) digest. Since this digest is computed over the entire IssuerSignedItemBytes (including the salt), data element contents cannot be inferred from digest values.

\textbf{Stage 2: MSO construction.} Aggregate all data element digests into valueDigests and construct the MobileSecurityObject structure together with digestAlgorithm, deviceKeyInfo, docType, and validityInfo. CBOR-encode the MSO and generate MobileSecurityObjectBytes as a tagged bytestring (CBOR tag 24).

\textbf{Stage 3: COSE\_Sign1 signing.} Based on the COSE\_Sign1 structure \cite{ref24}, generate a signature with MobileSecurityObjectBytes as the payload. The protected header includes the alg parameter (e.g., ES256 = -7), and the Sig\_Structure is constructed. Generate an ECDSA (P-256/P-384/P-521) or EdDSA signature over the Sig\_Structure using the issuer's private key. ISO/IEC 18013-5 specifies four algorithms: ES256, ES384, ES512, and EdDSA. The signer's public key certificate is stored in the COSE unprotected header's x5chain parameter.

\textbf{Stage 4: mdoc assembly.} The COSE\_Sign1 structure (= {[}protected\_header, unprotected\_header, payload, signature{]}) becomes the issuerAuth, and the array of IssuerSignedItemBytes becomes the nameSpaces, assembling the Document structure. The final mdoc is serialized as CBOR binary.

\subsubsection{Verification Process}

mdoc verification (ISO/IEC 18013-5 \S9.3.1 Inspection procedure for issuer data authentication) follows the procedure below.

\textbf{Stage 1: CBOR decoding and structure expansion.} CBOR-decode the received mdoc binary to obtain the Document structure. Expand the issuerAuth (COSE\_Sign1) within issuerSigned to obtain the protected header, payload (MobileSecurityObjectBytes), and signature value.

\textbf{Stage 2: COSE\_Sign1 signature verification.} Obtain the alg parameter from the protected header and confirm it is a permitted algorithm (ES256/ES384/ES512/EdDSA). Reconstruct the same Sig\_Structure as during signing, obtain the issuer's public key certificate from the COSE unprotected header's x5chain, verify the certificate chain (path validation to trust anchor, revocation check), and then verify the signature using the public key. Since the protected header is included in the Sig\_Structure and covered by the signature, protected header tampering is detected as a signature mismatch.

\textbf{Stage 3: Per-element digest verification.} Decode the MSO and obtain the digest map within valueDigests. Compute SHA-256 digests for each received IssuerSignedItemBytes and match them against the corresponding digest values in the MSO (linked by digestID). Data authenticity is confirmed only when all element digests match. This per-element digest method allows the verifier to individually verify the authenticity of only the received elements without needing all data elements.

\subsubsection{Selective Disclosure and Privacy Protection}

Selective disclosure in mdoc follows from its data model. While the MSO contains digests of all data elements, the holder sends only the data elements (IssuerSignedItems) to be disclosed to the verifier. The verifier can verify the authenticity of only the disclosed data elements by matching the digest of each received IssuerSignedItem against the corresponding digest in the MSO. For undisclosed data elements, digest values exist in the MSO, but the presence of the random field (salt of at least 16 bytes) makes it computationally infeasible to infer data element contents from digest values.

Furthermore, ISO/IEC 18013-5 specifies two methods of mdoc authentication: MAC authentication using DeviceKey and ECDSA/EdDSA signature authentication. MAC authentication uses an ephemeral MAC key (EMacKey) based on ECDH key agreement, where the holder's device computes the MAC. This method provides superior privacy protection compared to signature authentication, as the device can perform deniable authentication (since the verifier can also generate the MAC, device involvement cannot be proven to third parties).

\subsubsection{Extension to Online Presentation via ISO/IEC TS 18013-7}

ISO/IEC TS 18013-7 is a technical specification defining the online presentation protocol for mdoc, specifying the mdoc presentation flow via web browser using OID4VP (OpenID for Verifiable Presentations) \cite{ref27}. This enables remote mdoc verification in addition to the face-to-face presentation via NFC/BLE envisioned by ISO/IEC 18013-5. Since the mdoc data model and issuer signature structure are shared with ISO/IEC 18013-5, the results of this paper serve as a reference for evaluating the credential structure and signature verification components in online presentation. However, the overall protocol performance and security of ISO/IEC TS 18013-7/OID4VP, which includes session encryption, reader authentication, nonce processing, and transport/API-dependent processing, require separate evaluation.

\section{Experimental Design}

\subsection{Execution Environment}\label{sec:env}

The benchmarks were conducted in the four environments of Table~\ref{tab:t4}.

Every benchmark reported here---end-to-end issuance and verification, the cross-language comparison, serialization, attribute scaling, selective disclosure, the algorithm-unified benchmarks, the cryptographic primitive baseline and complex-credential canonicalization---was executed in all four, from one measurement session per environment. The four cover the hardware a verifier is plausibly deployed on: a cloud VM (A), a low-power bare-metal x86\_64 machine (B), a current desktop-class arm64 machine (C), and a server-class x86\_64 machine of around 2010 without AVX (D). Each session produced the same 162 benchmark conditions, with no condition missing from any environment, so every table in this paper reports the same comparison four times. Reporting each environment from a single session is deliberate: every figure attributed to an environment comes from the same session as every other figure for that environment, so the tables can be read against one another.

Three properties of this design are what let the four be read against each other. First, the measurement code is the same in all four. Second, the library versions are identical in all four (Table~\ref{tab:t5}). Third, all four run the same Node.js build and the same Go toolchain; the only version difference left is Python, 3.12.3 in Environment~A against 3.12.15 in B, C and D, so a difference between any two environments isolates the hardware.

\begin{table*}[!t]
\centering
\caption{The four execution environments. Every benchmark in this paper was run in all four. All four environments run the same Node.js build (v24.18.0, V8 13.6, OpenSSL 3.5.7), the same Go toolchain (go1.22.12) and the same library versions (Table~\ref{tab:t5}); Python is 3.12.3 in Environment~A and 3.12.15 in the other three, the only version difference left. The software is therefore held fixed and a difference between any two environments is attributable to the hardware (Section~\ref{sec:ttv}). Values quoted in the text without an environment are from Environment~A. Common to all four: high-resolution monotonic clocks (\texttt{process.hrtime.bigint()}, \texttt{time.Now()}, \texttt{time.perf\_counter\_ns()}); N=2,000 per benchmark after 50 warmup iterations; five independent process runs per benchmark, of which the median of each statistic is reported. Environments~A, B and C do not pin the measurement process to a core: on the cloud VM of Environment~A pinning did not reduce the run-to-run spread (Section~\ref{sec:env}), on Environment~B the clock is instead held at its 2.0\,GHz base by fixing the governor and disabling turbo, and macOS offers no equivalent of \texttt{taskset}.}\label{tab:t4}
\footnotesize\setlength{\tabcolsep}{4pt}
\begin{tabular}{@{}l@{\hspace{6pt}}>{\raggedright\arraybackslash}p{0.235\textwidth}@{\hspace{6pt}}>{\raggedright\arraybackslash}p{0.135\textwidth}@{\hspace{6pt}}>{\raggedright\arraybackslash}p{0.150\textwidth}@{\hspace{6pt}}>{\raggedright\arraybackslash}p{0.185\textwidth}@{}}
\toprule
\textbf{Env.} & \textbf{Hardware} & \textbf{OS / kernel} & \textbf{Runtimes} & \textbf{Isolation}\tabularnewline
\midrule
A & Cloud VM reserved for measurement (Azure, AMD EPYC 7763, x86\_64); not a physically dedicated host & Ubuntu Linux, 7.0.0-1017-azure & Node.js v24.18.0, Go 1.22.12, Python 3.12.3 & 2 of 4 vCPUs taken offline, SMT disabled; no core pinning\tabularnewline
\addlinespace
B & Bare metal, no virtualization (Intel Celeron N5095 @ 2.00\,GHz, x86\_64) & Linux 7.0.0-38-generic & Node.js v24.18.0, Go 1.22.12, Python 3.12.15 & SMT not supported by the CPU, governor \texttt{performance}, turbo disabled (base clock 2.0\,GHz); no core pinning\tabularnewline
\addlinespace
C & Apple M1 Ultra (arm64, 20 cores), desktop class & macOS (Darwin 27.0.0) & Node.js v24.18.0, Go 1.22.12, Python 3.12.15 & No core pinning (macOS provides no equivalent); otherwise idle\tabularnewline
\addlinespace
D & Intel Xeon E5645 @ 2.40\,GHz (Westmere-EP, 2 sockets $\times$ 6 cores, x86\_64; AES-NI but no AVX, AVX2 or SHA-NI) & Debian 13, Linux 6.12.107 & Node.js v24.18.0, Go 1.22.12, Python 3.12.15 & Pinned with \texttt{numactl} and \texttt{taskset} to one core of one NUMA node; SMT, turbo and frequency scaling disabled\tabularnewline
\bottomrule
\end{tabular}
\end{table*}

No measurement involved a GUI or a browser. Environment~D pins the measurement process to a single core of one NUMA node with \texttt{numactl}, binding memory to the same node, and has SMT, turbo and frequency scaling disabled; the SMT siblings are offline and the second socket is online but unused. Environments~A, B and C do not pin: on A, 2 of 4 vCPUs are offline and SMT is disabled; on B the governor is fixed to \texttt{performance} with turbo disabled, so the clock is held at the 2.0\,GHz base; on C, macOS offers no equivalent of \texttt{taskset}. On Environment~A, which is a cloud VM with two online vCPUs, we measured both ways and pinning did not help: over the conditions above 0.05\,ms, the run-to-run spread of the session reported here has a median of 1.6\% and a maximum of 12.7\% unpinned, against 5.6\% and 18.6\% pinned, and the number of conditions whose spread exceeds the 10\% band is one unpinned against three pinned. We read this as the hypervisor being free to take the one vCPU the process is pinned to, whereas an unpinned process can be moved; whatever the mechanism, we report the unpinned session because it is the quieter of the two. In every environment the suite was executed as five independent process runs per benchmark, and the median of each statistic (mean, $\sigma$, p50, p95) across those runs is reported.

Because only the run-to-run spread is a valid basis for comparing conditions---iterations within a run are autocorrelated through JIT compilation and garbage collection and are not independent samples---the end-to-end table reports, alongside p50, the spread of the five per-run p50 values as a percentage of the reported median. Across the end-to-end totals that spread is 0.3--1.6\% in Environment~A, 1.1--2.1\% in B, 2.0--4.3\% in C and 1.2--13.9\% in D. We accordingly treat two conditions as indistinguishable when their p50 values differ by less than 10\%, and say so wherever the judgment is used. One end-to-end condition exceeds that band on its own: the SD-JWT VC total in Environment~D, whose five runs span 13.9\%, which is exactly why we do not separate the mdoc and SD-JWT VC totals in that environment (Section~\ref{sec:e2e}). No other end-to-end total in any environment exceeds it. The per-iteration $\sigma$ and percentiles that the tables also carry describe the distribution within a run and are not used to support claims about differences between conditions.

Which statistics each environment contributes follows one rule throughout: all four environments carry p50, because that is the quantity conditions are compared on. Environment~A additionally carries the within-run distribution---mean and $\sigma$, with p95 as well in all but Table~\ref{tab:t10}---in Tables~\ref{tab:breakdown}, \ref{tab:micro}, \ref{tab:t10}, \ref{tab:t11} and \ref{tab:t13}, as an illustration of what that distribution looks like rather than because that environment has a privileged status; for the conditions in the remaining tables it is in the result files. Where the text quotes a figure without naming an environment, it is from Environment~A, and the surrounding sentence says what the other three do. Note that Environment~A is a cloud VM reserved for measurement rather than a physically dedicated host, so hypervisor-level effects cannot be fully excluded---the pinning result above is itself a symptom of that; Environments~B and D are bare metal and serve as the check on it (Section~\ref{sec:limits}).

\subsection{Test Tool Architecture}

This evaluation used a custom tool suite. All measurements reported in this paper were performed with a standalone CLI kit, vcbench \cite{ref28}; it consists of Node.js, Go, and Python engines plus a shared aggregation script, and it runs without any GUI or web server. Each engine records only raw per-iteration timings with high-resolution timestamps, and all statistics are computed centrally by the shared script, eliminating cross-language differences in statistical methodology. The source code is published in a GitHub repository \cite{ref28}, enabling replication. The result files record the iteration count actually completed for every condition. Every benchmark checks the result of the operation it times and raises if verification fails, so a silently failing verification cannot be reported as a fast one. Signing and verification for each format are implemented with the libraries listed in Table~\ref{tab:t5}.

\begin{table*}[!t]
\centering
\caption{Libraries used for each format (Node.js versions resolved from \texttt{node\_modules} at measurement time).}\label{tab:t5}
\footnotesize\setlength{\tabcolsep}{4pt}
\begin{tabular}{@{}>{\raggedright\arraybackslash}p{0.121\textwidth}>{\raggedright\arraybackslash}p{0.339\textwidth}>{\raggedright\arraybackslash}p{0.460\textwidth}@{}}
\toprule
\textbf{Format} & \textbf{Primary Library} & \textbf{Role}\tabularnewline
\midrule
SD-JWT VC & node:crypto (Node.js v24.18.0 / OpenSSL 3.5.7) & EdDSA (Ed25519) signing/verification\tabularnewline
SD-JWT VC & jose@6.2.8 & Reference implementation for JWT structure processing (complexity analysis and reference measurements)\tabularnewline
W3C VCDM & jsonld@8.3.3 (rdf-canonize@3.4.0) & JSON-LD processing / RDFC-1.0 canonicalization\tabularnewline
W3C VCDM & node:crypto (Ed25519, SHA-256) & Ed25519 signing / SHA-256 hashing\tabularnewline
W3C VCDM (JCS) & canonicalize@1.0.8 & JCS (RFC 8785) canonicalization\tabularnewline
mdoc & cbor-x@1.6.5 & CBOR encoding/decoding\tabularnewline
mdoc & node:crypto (ECDSA P-256) & ECDSA P-256 signing / COSE\_Sign1 verification\tabularnewline
W3C VCDM (ecdsa-sd-2023) & @digitalbazaar/ecdsa-sd-2023-cryptosuite@3.4.1, @digitalbazaar/data-integrity@2.5.0, jsonld-signatures@11.6.0 & Selective disclosure that preserves the issuer signature (Section~\ref{sec:sd})\tabularnewline
(Go engine) & github.com/piprate/json-gold v0.8.0, github.com/fxamacker/cbor/v2 v2.9.2 & RDFC-1.0 canonicalization / CBOR in Go\tabularnewline
(Python engine) & cryptography 50.0.2, PyLD 3.3.0, cbor2 6.1.5 & Signatures / RDFC-1.0 canonicalization / CBOR in Python\tabularnewline
(Complex credential evaluation) & @digitalcredentials/open-badges-context@3.0.0, @digitalbazaar/credentials-context@3.2.0 & Statically embedded Open Badges v3.0 / VCDM 2.0 contexts (Section~\ref{sec:complex})\tabularnewline
\bottomrule
\end{tabular}

\vspace{2pt}
{\footnotesize The jsonld library exposes this canonicalization under the algorithm identifier \texttt{URDNA2015}; that identifier therefore appears in the code listings, and denotes the same algorithm as RDFC-1.0.}
\end{table*}

\subsection{Benchmarking Methodology}

\subsubsection{Signature Verification Speed Benchmark}\label{sec:bench-speed}

For each format, signing (sign) and verification (verify) were repeatedly executed on identical credential information (name, date of birth, address, etc.), measuring the time per operation (ms/op; ops/s is its reciprocal). The scope covered by the main measurements differs per format and is stated here explicitly. For SD-JWT VC it covers construction of the signing input and JWS signature generation, and on the verifying side signature verification and the check of required claims; it does not include creation of Disclosures or matching of their digests. For VCDM 2.0 + Data Integrity it covers the full chain of RDFC-1.0 canonicalization, SHA-256 hashing, and Ed25519 signing. The simple credential used in these measurements has five quads and one blank node, and its \texttt{@context} is inline: it defines only the terms the credential uses and references no external context document, so these figures include no context-document expansion, which a credential carrying \texttt{https://www.w3.org/ns/credentials/v2} would add (Section~\ref{sec:complex} measures that term on production schemas). For mdoc it covers CBOR encoding, MSO construction with a SHA-256 digest per data element, and COSE\_Sign1 signing, and the verifying side also includes per-element digest matching; the verifier holds the issuer key directly, so x5chain path validation is excluded. In other words, the main measurements target the core signing and signature-verification path of a single credential, and the cost specific to selective disclosure falls outside that path for SD-JWT VC only. This paper therefore presents, in Table~\ref{tab:e2e} of Section~\ref{sec:e2e}, only the end-to-end measurement in which all three formats cover an identical scope (issue, present with selective disclosure, and verify the disclosed claims) as the performance comparison, and treats the partial paths---signing, verification, canonicalization---as a decomposition (Sections~\ref{sec:breakdown} and \ref{sec:lang}). Section~\ref{sec:seldisc} breaks the cost down by the number of disclosed attributes, and the point is revisited in the construct validity discussion of Section~\ref{sec:ttv}. Measurements used monotonic clocks with nanosecond resolution (Node.js: process.hrtime.bigint(), Go: time.Now(), Python: time.perf\_counter\_ns()), so that per-iteration durations in the microsecond range are resolved, recording the duration of each iteration individually. Fifty warmup iterations preceded the main measurement to reduce bias from JIT optimization and cache initialization. The statistical design---separating warmup, repeating independent runs, and reporting medians---follows the methodology of Georges et al. \cite{ref29} for performance evaluation under the non-determinism introduced by JIT compilation and garbage collection. These timers have far finer resolution than the browser's performance.now() (DOMHighResTimeStamp, whose resolution is limited to roughly 0.1\,ms by Spectre mitigations) and are suitable for measuring sub-millisecond operations. This experiment computed the arithmetic mean, sample standard deviation ($\sigma$), 95\% confidence interval (95\%CI, in the result files), median (p50), and 95th percentile (p95) over N=2,000 iterations per benchmark; all statistics, including p50 and p95, are reported to three decimal places. Percentiles are computed with linear interpolation. Ratios and percentages quoted in the text are computed from the unrounded medians, so recomputing them from the three-digit values in the tables can differ in the last digit shown.

Regarding outliers, values beyond the Tukey fences (1.5$\times$IQR, the interquartile range, from the first/third quartiles) are detected and reported but are not removed from the data. Because the per-iteration distributions are right-skewed due to V8 garbage collection and OS scheduling, the median (p50), which is robust to outliers, serves as the primary statistic, with the mean reported for reference. Outlier rates in the main benchmarks range from about 4\% to 20\%, and the Tukey-trimmed means closely match the medians. In addition, the entire suite was repeated as five independent process runs, and the median of each statistic across the runs is reported, mitigating systematic noise (e.g., interference from other processes) affecting any single run. Note that the $\sigma$ reported in the tables describes the spread within a single run, not the spread across runs; run-to-run spread is reported as the range of the five per-run p50 values. Over the conditions whose p50 exceeds 0.05\,ms---below that, a percentage of a sub-50-microsecond median is itself noise---that spread is 0.3--12.7\% in Environment~A, 0.0--23.0\% in B, 0.7--9.8\% in C and 0.0--17.0\% in D; for the end-to-end totals specifically it is 0.3--1.6\%, 1.1--2.1\%, 2.0--4.3\% and 1.2--13.9\%. For the slowest conditions (W3C VCDM at 100 and 500 attributes in Table~\ref{tab:t11}, and OB3, DCC-style and 50-blank-node credentials in Table~\ref{tab:t14}) the iteration count was reduced (N=400 for the scaling conditions, N=200--1,000 for the complex credentials, and N=100 for the \texttt{ecdsa-sd-2023} conditions of Table~\ref{tab:sd}); for those conditions we confirmed that the run-to-run variation of p50 remained comparable before reporting them.

The core logic of the benchmark is shown in the following pseudocode. Two versions are provided---synchronous (bench) and asynchronous (benchAsync)---which call each format's sign/verify function N times, recording each iteration's duration individually to compute distribution statistics.

\Needspace*{5\baselineskip}
\textbf{Listing 6. Core Logic of the Benchmark Helper (bench / benchAsync, pseudocode)}\nopagebreak

\begin{lstlisting}
// Backend Benchmark (Node.js)
function bench(label, n, fn) {
  for (let i = 0; i < 50; i++) fn(); // warmup x50
  const t = new Array(n);
  for (let i = 0; i < n; i++) {
    const s = process.hrtime.bigint();
    fn();
    t[i] = Number(process.hrtime.bigint() - s); // record each iteration
  }
  return stats(t); // mean, sigma, 95%CI, p50, p95, outliers (Tukey 1.5xIQR)
}
\end{lstlisting}

For W3C VCDM, in addition to the overall signing process, internal steps (RDFC-1.0 canonicalization, SHA-256 hashing, Ed25519 signing) were individually measured using process.hrtime.bigint(). Timestamps were captured before and after each step, and their difference was recorded as the duration of each step.

\subsubsection{Complexity of the Verification Reference Implementations}\label{sec:bench-cplx}

For the verification implementation of each format, five static and dynamic metrics were measured: lines of code (LOC), number of asynchronous steps, cyclomatic complexity \cite{ref30}, number of external network calls, and number of external dependency libraries. These metrics serve as indirect indicators of implementation maintainability, auditability, and vulnerability introduction risk. Note that LOC depends on implementation style (degree of error handling omission, library utilization), and cyclomatic complexity is calculated for the entire deserialization function. The number of external dependency libraries is a proxy indicator for supply chain risk, but individual library maturity, maintenance status, and CVE (Common Vulnerabilities and Exposures) history were excluded from evaluation. These limitations are discussed in Threats to Validity (Section~\ref{sec:ttv}).

\subsubsection{Canonicalization Evaluation on Complex Credentials}\label{sec:bench-canon}

The credentials used in the main benchmark (Section~\ref{sec:bench-speed}) have a simple structure with a single blank node and may therefore underestimate RDFC-1.0 canonicalization cost. To address this, we measured RDFC-1.0 canonicalization time on complex, production-grade schemas (Section~\ref{sec:complex}): a 1EdTech Open Badges v3.0 (OB3) AchievementCredential, widely adopted in the education domain, and a sample modeled on the OB3-profile academic degree credentials issued by the Digital Credentials Commons (DCC). OB3 credentials contain multiple nodes without \texttt{id}s (blank nodes in the RDF dataset) under \texttt{achievement}---\texttt{criteria}, \texttt{alignment}, and \texttt{result}---and OB3 is also the representative complex schema used in VC interoperability efforts such as the JFF Plugfest \cite{ref31}, a series of events hosted by Jobs for the Future (JFF) together with the W3C Verifiable Credentials for Education Task Force, in which implementers test their issuers, wallets and verifiers against one another. The JSON-LD contexts (VCDM 2.0 and OB3 v3.0.3) were statically embedded from the \texttt{@digitalcredentials/open-badges-context} and \texttt{@digitalbazaar/credentials-context} packages, eliminating network I/O during measurement. In addition, synthetic credentials containing 10 and 50 id-less child nodes were measured to isolate the effect of blank node count.

The implementation code used for the measurements, together with the reference implementations analyzed in the complexity study, is provided in Appendix A.

\section{Experimental Results}

\subsection{End-to-End Performance with Selective Disclosure}\label{sec:e2e}

Table~\ref{tab:e2e} reports the basis for our performance comparison: a measurement in which all three formats cover an identical scope. The measured scenario is issuing a credential with 20 attributes, disclosing 5 of them, and verifying up to the disclosed claims. A comparison restricted to signing and verification alone would not be like for like, because only for SD-JWT VC would the creation and digest matching of Disclosures fall outside the measured path. Within this scope, SD-JWT VC creates the Disclosures, stores their digests in the \texttt{\_sd} array, selects them at presentation time and matches digests at verification; W3C VCDM canonicalizes the document and the proof options, hashes both and signs the 64-byte concatenation, as \texttt{eddsa-rdfc-2022} requires; and mdoc builds an MSO with a SHA-256 digest and a 16-byte salt per element, and at verification decodes the CBOR, checks COSE\_Sign1 and matches each disclosed element's digest against the MSO by \texttt{digestID}.

One asymmetry in that scope has to be stated before the numbers, because it is not a difference in speed but in what the verifier ends up knowing. SD-JWT VC and mdoc both let the holder reveal a subset while the \emph{issuer's} signature still covers the revealed values. \texttt{eddsa-rdfc-2022} has no derivation step, so the only way to present a subset is to canonicalize and sign it again---and the signature the verifier then checks is the holder's, not the issuer's. The note beneath Table~\ref{tab:e2e} records this. Section~\ref{sec:sd} measures \texttt{ecdsa-sd-2023}, the Data Integrity cryptosuite that does preserve the issuer signature, and that is the row to read when the comparison needs to hold the guarantee fixed rather than the scope.

\begin{table*}[!t]
\centering
\caption{End-to-end performance with selective disclosure (5 of 20 attributes disclosed; issue $\rightarrow$ present $\rightarrow$ verify, p50 in ms, medians across N=2,000 $\times$ 5 runs), in all four environments. \emph{spread} is the run-to-run spread of the five per-run p50 values of the total. The ordering mdoc $<$ SD-JWT VC $<$ W3C VCDM holds in all four; in D, however, the mdoc and SD-JWT VC totals are only 10\% apart while the SD-JWT VC total's own run-to-run spread is 13.9\%, so that environment does not separate those two. The VP size is the serialized form sent to the verifier and is identical in all four environments. The three scopes are not equivalent in what the verifier learns, which the note beneath the table states and which Section~\ref{sec:sd} measures at equal guarantee.}\label{tab:e2e}
\footnotesize\setlength{\tabcolsep}{5pt}
\begin{tabular}{@{}l r r r r r@{}}
\toprule
& \multicolumn{4}{c}{\textbf{p50 (ms)}} & \tabularnewline
\cmidrule(lr){2-5}
\textbf{Format / environment} & \textbf{issue} & \textbf{present} & \textbf{verify} & \textbf{total} & \textbf{spread (\%)}\tabularnewline
\midrule
\multicolumn{6}{@{}l}{\textit{SD-JWT VC} (VP 1{,}878 bytes)}\tabularnewline
\quad A (EPYC 7763) & 0.147 & 0.000 & 0.141 & 0.298 & 0.3\tabularnewline
\quad B (Celeron N5095) & 0.445 & 0.001 & 0.420 & 0.875 & 2.1\tabularnewline
\quad C (Apple M1 Ultra) & 0.078 & 0.000 & 0.095 & 0.179 & 3.1\tabularnewline
\quad D (Xeon E5645) & 0.565 & 0.001 & 0.315 & 0.818 & 13.9\tabularnewline
\addlinespace
\multicolumn{6}{@{}l}{\textit{W3C VCDM} (VP 1{,}549 bytes)}\tabularnewline
\quad A (EPYC 7763) & 0.191 & 0.134 & 0.212 & 0.534 & 0.7\tabularnewline
\quad B (Celeron N5095) & 0.496 & 0.350 & 0.583 & 1.390 & 1.1\tabularnewline
\quad C (Apple M1 Ultra) & 0.096 & 0.069 & 0.125 & 0.287 & 2.0\tabularnewline
\quad D (Xeon E5645) & 0.482 & 0.354 & 0.485 & 1.287 & 1.6\tabularnewline
\addlinespace
\multicolumn{6}{@{}l}{\textit{mdoc} (VP 1{,}561 bytes)}\tabularnewline
\quad A (EPYC 7763) & 0.122 & 0.004 & 0.128 & 0.265 & 1.6\tabularnewline
\quad B (Celeron N5095) & 0.303 & 0.011 & 0.358 & 0.682 & 2.1\tabularnewline
\quad C (Apple M1 Ultra) & 0.067 & 0.003 & 0.081 & 0.150 & 4.3\tabularnewline
\quad D (Xeon E5645) & 0.355 & 0.010 & 0.377 & 0.744 & 1.2\tabularnewline
\bottomrule
\end{tabular}

\vspace{2pt}
{\footnotesize What the verifier checks: for SD-JWT VC, the issuer signature over the disclosed values; for W3C VCDM, the \textbf{holder's} signature over a re-signed subset, not the issuer's; for mdoc, the issuer signature over the MSO, plus a per-element digest match.}
\end{table*}

The totals order the same way in all four environments: mdoc lowest, SD-JWT VC next, W3C VCDM last. The absolute values span about a factor of five between the fastest and slowest machine---mdoc's total is 0.150\,ms in Environment~C and 0.744\,ms in D---so the result that transfers is the ordering and the structure beneath it, not the numbers.

What the four environments agree on, they agree on by a margin. W3C VCDM's total is 1.73--2.04$\times$ mdoc's and 1.57--1.79$\times$ SD-JWT VC's, and the narrowest of those eight ratios is five times the indistinguishability band of Section~\ref{sec:env}. What they do not agree on is the gap between the other two: mdoc and SD-JWT VC differ by 12\% in Environment~A, 28\% in B, 19\% in C and 10\% in D. The smallest of these sits exactly at the edge of the band, and the SD-JWT VC total in that environment has a run-to-run spread of 13.9\% on its own---larger than the gap it would have to resolve---so Environment~D does not separate mdoc from SD-JWT VC at all. A deployment choosing between those two on throughput grounds therefore cannot read the decision off this table without knowing its own hardware; a deployment weighing either against W3C VCDM can.

The breakdown locates the W3C VCDM gap, and locates it in the same place in each environment. In Environment~A verification costs 0.128\,ms for mdoc, 0.141\,ms for SD-JWT VC and 0.212\,ms for W3C VCDM. Presentation is below one microsecond for SD-JWT VC, which only concatenates the selected Disclosures, and 4\,$\mu$s for mdoc, which slices the disclosed elements out of an already-signed structure; W3C VCDM needs 0.134\,ms because it canonicalizes and signs the subset from scratch. Issuance falls between 0.12 and 0.20\,ms for all three. The same shape appears in B, C and D: presentation is microseconds for SD-JWT VC and mdoc and two orders of magnitude more for W3C VCDM (0.350, 0.069 and 0.354\,ms respectively), and the presentation step alone accounts for about a quarter of W3C VCDM's total in every environment (24.0--27.5\%). The single structural reason is that RDFC-1.0 canonicalization runs in all three steps---issuance signs the credential, presentation re-signs the disclosed subset, and verification recomputes the hash---and runs twice within each, once over the document and once over the proof options, so one end-to-end exchange canonicalizes six times.

One step does change its ordering with the machine, and it is worth isolating. Verification is cheaper for mdoc than for SD-JWT VC in Environments~A, B and C (0.128 against 0.141, 0.358 against 0.420, 0.081 against 0.095\,ms) and more expensive in D (0.377 against 0.315\,ms). Environment~D is the machine without AVX, where Node.js verifies P-256 more slowly than Ed25519 while the other three do the reverse (Table~\ref{tab:prim}; P-256 signing is the slower of the two in Node.js everywhere), and mdoc is the only one of the three formats using P-256 in its default configuration. The reversal is thus a property of the primitive on that hardware rather than of the mdoc pipeline, which Section~\ref{sec:unified} confirms by holding the algorithm fixed.

\subsection{Separating the Algorithm from the Pipeline}\label{sec:unified}

The totals above pair each format with the signature algorithm its deployments use, which leaves the comparison confounded: mdoc runs on ECDSA P-256 while the other two run on Ed25519. Table~\ref{tab:unified-e2e} repeats the identical scenario twice more, once with Ed25519 for all three formats and once with P-256 for all three (SD-JWT VC on ES256, Data Integrity on \texttt{ecdsa-rdfc-2019}).

\begin{table*}[!t]
\centering
\caption{The scenario of Table~\ref{tab:e2e} with one signature algorithm across all three formats, in all four environments (p50 in ms, medians across N=2,000 $\times$ 5 runs). Unlike Table~\ref{tab:t13}, this table measures the same scope as Table~\ref{tab:e2e}, so the figures in the two tables are comparable; the rows that repeat a configuration of Table~\ref{tab:e2e} are a separate measurement and differ from it by a few microseconds. In the P-256 block SD-JWT VC is on ES256 and Data Integrity on \texttt{ecdsa-rdfc-2019}. mdoc on Ed25519 (COSE alg $-8$) is a configuration ISO/IEC 18013-5 permits but deployed mDL profiles do not use by default, and is measured only to separate the algorithm from the pipeline.}\label{tab:unified-e2e}
\footnotesize\setlength{\tabcolsep}{5pt}
\begin{tabular}{@{}l c r r r r@{}}
\toprule
& & \multicolumn{4}{c}{\textbf{p50 (ms)}}\tabularnewline
\cmidrule(lr){3-6}
\textbf{Format} & \textbf{Env.} & \textbf{issue} & \textbf{present} & \textbf{verify} & \textbf{total}\tabularnewline
\midrule
\multicolumn{6}{@{}l}{\textit{Ed25519 for all three (mdoc on COSE alg $-8$)}}\tabularnewline
SD-JWT VC & A & 0.148 & 0.000 & 0.142 & 0.298\tabularnewline
 & B & 0.445 & 0.001 & 0.418 & 0.874\tabularnewline
 & C & 0.079 & 0.000 & 0.095 & 0.175\tabularnewline
 & D & 0.572 & 0.001 & 0.313 & 0.812\tabularnewline
\addlinespace
W3C VCDM & A & 0.190 & 0.134 & 0.213 & 0.537\tabularnewline
 & B & 0.499 & 0.351 & 0.581 & 1.386\tabularnewline
 & C & 0.096 & 0.070 & 0.125 & 0.286\tabularnewline
 & D & 0.483 & 0.354 & 0.487 & 1.291\tabularnewline
\addlinespace
mdoc & A & 0.102 & 0.004 & 0.140 & 0.253\tabularnewline
 & B & 0.282 & 0.010 & 0.414 & 0.708\tabularnewline
 & C & 0.066 & 0.003 & 0.096 & 0.164\tabularnewline
 & D & 0.286 & 0.010 & 0.308 & 0.608\tabularnewline
\midrule
\multicolumn{6}{@{}l}{\textit{ECDSA P-256 for all three}}\tabularnewline
SD-JWT VC & A & 0.164 & 0.000 & 0.127 & 0.301\tabularnewline
 & B & 0.465 & 0.001 & 0.355 & 0.838\tabularnewline
 & C & 0.080 & 0.000 & 0.079 & 0.160\tabularnewline
 & D & 0.647 & 0.001 & 0.381 & 0.948\tabularnewline
\addlinespace
W3C VCDM & A & 0.225 & 0.166 & 0.207 & 0.602\tabularnewline
 & B & 0.568 & 0.418 & 0.553 & 1.496\tabularnewline
 & C & 0.098 & 0.071 & 0.107 & 0.268\tabularnewline
 & D & 0.554 & 0.424 & 0.551 & 1.500\tabularnewline
\addlinespace
mdoc & A & 0.122 & 0.004 & 0.127 & 0.264\tabularnewline
 & B & 0.303 & 0.011 & 0.357 & 0.678\tabularnewline
 & C & 0.067 & 0.003 & 0.081 & 0.152\tabularnewline
 & D & 0.353 & 0.010 & 0.376 & 0.742\tabularnewline
\bottomrule
\end{tabular}
\end{table*}

The ordering survives both substitutions, in all four environments: mdoc is lowest, SD-JWT VC next, W3C VCDM last. The headline result is therefore a property of the pipelines rather than of the algorithm pairing. What the algorithm does change is the composition of the mdoc total. Under Ed25519 mdoc's verification becomes 0.140\,ms against SD-JWT VC's 0.142\,ms---the two are indistinguishable, so mdoc is no longer the faster verifier---and it keeps the lower total only because its issuance is cheaper (0.102 against 0.148\,ms). The claim that mdoc's advantage rests on its P-256/COSE configuration is thus correct about the verification step and incorrect about the total, and Section~\ref{sec:lang} shows why: which of Ed25519 and P-256 verifies faster depends on the runtime and the machine.

\subsection{Selective Disclosure That Preserves the Issuer Signature}\label{sec:sd}

Table~\ref{tab:e2e} measures what an \texttt{eddsa-rdfc-2022} deployment can actually do, which is to re-sign the subset. To compare mechanisms that offer SD-JWT VC's and mdoc's guarantee, Table~\ref{tab:sd} measures \texttt{ecdsa-sd-2023} \cite{ref22}, the Data Integrity cryptosuite in which the issuer signs once and the holder derives a presentation that still carries that signature.

\begin{table}[!t]
\centering
\caption{\texttt{ecdsa-sd-2023} on P-256: selective disclosure preserving the issuer signature, in all four environments (p50 in ms, medians across 5 runs, N=100 per condition). Issuance cost does not depend on how many attributes are disclosed later.}\label{tab:sd}
\footnotesize\setlength{\tabcolsep}{3pt}
\begin{tabular}{@{}l r r r r r@{}}
\toprule
& \multicolumn{4}{c}{\textbf{p50 (ms)}} & \tabularnewline
\cmidrule(lr){2-5}
\textbf{Operation} & \textbf{A} & \textbf{B} & \textbf{C} & \textbf{D} & \textbf{Bytes}\tabularnewline
\midrule
Issue (base proof, 20 attrs) & 3.367 & 7.398 & 1.528 & 10.254 & 2{,}917\tabularnewline
Derive 1 of 20 & 1.871 & 4.887 & 0.979 & 4.493 & 802\tabularnewline
Derive 2 of 20 & 1.795 & 4.560 & 0.956 & 4.048 & 913\tabularnewline
Derive 5 of 20 & 1.807 & 4.359 & 0.910 & 3.951 & 1{,}246\tabularnewline
Derive 10 of 20 & 1.833 & 4.465 & 0.924 & 4.039 & 1{,}801\tabularnewline
Derive 20 of 20 & 1.949 & 4.681 & 0.992 & 4.274 & 2{,}911\tabularnewline
Verify 1 of 20 & 1.598 & 3.925 & 0.782 & 3.716 & ---\tabularnewline
Verify 2 of 20 & 1.512 & 3.629 & 0.718 & 3.575 & ---\tabularnewline
Verify 5 of 20 & 1.668 & 3.894 & 0.788 & 4.466 & ---\tabularnewline
Verify 10 of 20 & 2.027 & 4.330 & 0.887 & 5.963 & ---\tabularnewline
Verify 20 of 20 & 2.431 & 5.117 & 1.061 & 9.014 & ---\tabularnewline
\bottomrule
\end{tabular}
\end{table}

Holding the guarantee fixed changes the comparison by an order of magnitude, and does so in every environment. Issuing a base proof costs 3.367\,ms in Environment~A, against 0.191\,ms for the \texttt{eddsa-rdfc-2022} path and 0.147\,ms for SD-JWT VC; the ratio of base-proof issuance to \texttt{eddsa-rdfc-2022} issuance is 17.6$\times$ in A, 14.9$\times$ in B, 15.9$\times$ in C and 21.3$\times$ in D. Deriving a 5-of-20 presentation costs 1.807\,ms in Environment~A against 0.134\,ms, and verifying it 1.668\,ms against 0.212\,ms for W3C VCDM and 0.141\,ms for SD-JWT VC; the same three comparisons hold with the same sign in B, C and D. The derived presentation is also the one whose verification cost grows with the number of attributes revealed---from 1.598\,ms at one attribute to 2.431\,ms at twenty in Environment~A---while derivation itself is nearly flat above two attributes everywhere. The size of that growth does not transfer: it is 1.5$\times$ in A, 1.3$\times$ in B, 1.4$\times$ in C and 2.4$\times$ in D, so we report the growth as present in all four and bounded by 1.3--2.4$\times$ rather than as a constant.

The reading we draw is narrow and worth stating precisely. On the Data Integrity side, a deployment chooses between a presentation that is cheap but drops the issuer's signature and one that keeps it at roughly ten times the cost; SD-JWT VC and mdoc provide the second property at the cost of the first. The 2.02$\times$ in Table~\ref{tab:e2e} is therefore a lower bound on what W3C VCDM costs relative to the other two when the comparison is made at equal guarantee, and we use it as such in Section~\ref{sec:disc}.

\subsection{Where the W3C VCDM Signing Cost Goes}\label{sec:breakdown}

Table~\ref{tab:breakdown} decomposes the W3C VCDM signing pipeline by measuring each stage on its own, under the conditions of Table~\ref{tab:micro} (the simple credential).

\begin{table*}[!t]
\centering
\caption{W3C VCDM signing pipeline, each stage measured individually (simple credential; all columns in ms, medians across N=2,000 $\times$ 5 runs). The \texttt{normalize} row is the cost of \emph{one} canonicalization; \texttt{eddsa-rdfc-2022} performs two, over the document and over the proof options. The mean, $\sigma$ and p95 columns describe the within-run distribution in Environment~A; p50 is given for all four environments, because that is the basis on which conditions are compared.}\label{tab:breakdown}
\footnotesize\setlength{\tabcolsep}{5pt}
\begin{tabular}{@{}l r r r r r r r@{}}
\toprule
& \multicolumn{4}{c}{\textbf{Env.~A (ms)}} & \multicolumn{3}{c}{\textbf{p50 (ms)}}\tabularnewline
\cmidrule(lr){2-5}\cmidrule(lr){6-8}
\textbf{Stage} & \textbf{mean} & \textbf{$\sigma$} & \textbf{p50} & \textbf{p95} & \textbf{B} & \textbf{C} & \textbf{D}\tabularnewline
\midrule
RDFC-1.0 canonicalization (one pass) & 0.053 & 0.043 & 0.042 & 0.092 & 0.119 & 0.020 & 0.156\tabularnewline
SHA-256 hashing & 0.003 & 0.015 & 0.002 & 0.004 & 0.006 & 0.001 & 0.010\tabularnewline
Ed25519 signing & 0.039 & 0.030 & 0.037 & 0.047 & 0.108 & 0.029 & 0.086\tabularnewline
\midrule
Pipeline with one canonicalization & 0.102 & 0.031 & 0.095 & 0.124 & 0.244 & 0.052 & 0.263\tabularnewline
Pipeline as the cryptosuite specifies it & 0.154 & 0.055 & 0.139 & 0.210 & 0.374 & 0.072 & 0.414\tabularnewline
\bottomrule
\end{tabular}
\end{table*}

Taking the stages at face value, one canonicalization (0.042\,ms in Environment~A) and the Ed25519 signature (0.037\,ms) are the same order of magnitude and SHA-256 hashing (0.002\,ms) is negligible. That reading understates canonicalization, because the cryptosuite canonicalizes twice. The last two rows show the difference directly: the pipeline costs 0.095\,ms with one canonicalization and 0.139\,ms as specified, and the 0.044\,ms gap matches the 0.042\,ms of a second pass. The same arithmetic closes in the other three environments, and the ratio of the two-canonicalization measurement to the one-canonicalization measurement, in the canonicalization-only rows of Table~\ref{tab:micro}, is 1.90, 1.91, 1.99 and 1.94 in A, B, C and D, which is the one quantity in this subsection that is the same everywhere---as it should be, since it is a property of the cryptosuite rather than of the machine.

Against the specified pipeline, canonicalization accounts for 61.0\% of signing in Environment~A, signing for 26.6\% and hashing for 1.3\%. These shares, measured against the same specified pipeline, move with the machine: canonicalization and signing take 63.4\% and 28.8\% in Environment~B, 56.6\% and 40.5\% in C, and 75.4\% and 20.8\% in D. That canonicalization in total is the largest stage holds in all four environments, but it does so because the cryptosuite runs it twice: counted as a single pass it is the larger stage in A, B and D (44.2\%, 48.8\% and 59.3\% against 38.9\%, 44.3\% and 32.7\% for signing) and the smaller one in C (38.5\% against 55.8\%). The share canonicalization takes does not transfer between machines, and a capacity estimate built from one machine's share will be wrong on another by up to a factor of two. Even for a credential with a single blank node, then, canonicalization is already the largest stage of the pipeline the cryptosuite specifies---and Section~\ref{sec:complex} shows that for production schemas it stops being merely the largest and becomes dominant.

\subsection{Structural Complexity of the Reference Implementations}\label{sec:cplx}

The metrics in Table~\ref{tab:t8} are counted over the verification wrappers written for this paper, not over the formats. We report them because the wrappers were written to the same brief---parse the credential, prepare the signing input, check the issuer signature---so the differences between them reflect how much work each specification places in that brief. They are not a measure of the inherent complexity of the formats, and the ordering they produce would change with a different decomposition into functions. Cyclomatic complexity is McCabe's count of decision points plus one, with loops counted \cite{ref32}; we state the definition because the measure has well-known limitations as a proxy for comprehension difficulty \cite{ref30}. Lines of code and the number of asynchronous steps are reported alongside it for that reason.

\begin{table}[!t]
\centering
\caption{Structural metrics of the verification reference implementations written for this paper (Appendix A.4). These describe our wrappers, not the formats; see the caveats in the text and in Section~\ref{sec:ttv}.}\label{tab:t8}
\footnotesize\setlength{\tabcolsep}{2pt}
\begin{tabular}{@{}l r r r@{}}
\toprule
\textbf{Metric} & \textbf{SD-JWT VC} & \textbf{W3C VCDM} & \textbf{mdoc}\tabularnewline
\midrule
Lines of code & 10 & 35 & 25\tabularnewline
Asynchronous steps & 1 & 4 & 2\tabularnewline
Cyclomatic complexity & 5 & 9 & 8\tabularnewline
Network calls, as measured & 0 & 0 & 0\tabularnewline
Network calls at defaults & 0 & 2 & 0\tabularnewline
Direct dependencies & 1 & 2 & 2\tabularnewline
\quad which & \texttt{jose} & \texttt{jsonld}, & \texttt{cbor-x},\tabularnewline
& & \texttt{node:crypto} & \texttt{node:crypto}\tabularnewline
\bottomrule
\end{tabular}
\end{table}

Two rows need comment. The network-call row is split because the two answers differ: the code we measured resolves every context from a statically embedded map and issues no request, while, for a credential that references its contexts by URL, the same code at the library's default document loader would fetch two contexts over HTTP (the measured credential embeds its context, so for it no fetch would occur even at defaults). The distinction matters for a deployment, which chooses between the two. The dependency counts are of direct imports in the wrapper; \texttt{node:crypto} is part of the Node.js runtime and is listed for completeness rather than as a supply-chain dependency.

The wrappers themselves (Listings 13--15, Appendix A.4) show where the counts come from. SD-JWT VC delegates to a single \texttt{jwtVerify} call, which is why ten lines suffice---and why the comparison flatters it, since the Disclosure matching that Section~\ref{sec:e2e} measures is not in the wrapper at all. W3C VCDM passes through context resolution, one canonicalization, hashing, signature verification and schema checks; this wrapper predates the specification-aligned benchmark of Table~\ref{tab:micro}, canonicalizes the document only, and the counts are of it as listed. mdoc separates CBOR decoding from COSE\_Sign1 verification and then loops over the disclosed elements, matching each one's digest against the MSO by \texttt{digestID}; the loop is what takes its McCabe count above SD-JWT VC's despite the smaller line count, and indexing by \texttt{digestID} rather than by position is what keeps it correct when only a subset is disclosed.

\subsection{Decomposing the Cost: Libraries, Specifications and Primitives}\label{sec:lang}

Section~\ref{sec:e2e} established the totals; this section asks what inside the pipelines produces them. Table~\ref{tab:micro} measures signing and verification of a single credential without selective disclosure, and Table~\ref{tab:prim} measures the signature primitives with no credential structure around them at all. The second table exists because the first cannot be interpreted without it: a format benchmark pairs an algorithm with a pipeline, and attributing a difference to one or the other requires knowing what the algorithm alone costs.

\begin{table*}[!t]
\centering
\caption{Signing and verification of a single credential, Node.js, in all four environments (p50 in ms, medians across N=2,000 $\times$ 5 runs). The mean, $\sigma$, p95 and spread columns describe Environment~A; p50 is given for all four. The W3C VCDM rows canonicalize twice, as \texttt{eddsa-rdfc-2022} requires, and the mdoc \texttt{cbor-x} rows decode CBOR and match each disclosed element's digest against the MSO, so the two formats are measured at the scope their specifications define. The supplementary block is not an alternative implementation of the same guarantee: \emph{canonicalization removed} signs a statically expanded N-Quads document and cannot establish that the signature covers what was received.}\label{tab:micro}
\footnotesize\setlength{\tabcolsep}{3pt}
\begin{tabular}{@{}l l l r r r r r r r r@{}}
\toprule
& & & \multicolumn{5}{c}{\textbf{Env.~A (ms; spread in \%)}} & \multicolumn{3}{c}{\textbf{p50 (ms)}}\tabularnewline
\cmidrule(lr){4-8}\cmidrule(lr){9-11}
\textbf{Format} & \textbf{Mode} & \textbf{Op} & \textbf{mean} & \textbf{$\sigma$} & \textbf{p50} & \textbf{p95} & \textbf{spread} & \textbf{B} & \textbf{C} & \textbf{D}\tabularnewline
\midrule
SD-JWT VC & node:crypto & sign & 0.041 & 0.020 & 0.038 & 0.050 & 0.2 & 0.113 & 0.030 & 0.094\tabularnewline
SD-JWT VC & node:crypto & verify & 0.119 & 0.012 & 0.117 & 0.130 & 3.3 & 0.350 & 0.086 & 0.248\tabularnewline
W3C VCDM & jsonld (as specified) & sign & 0.154 & 0.055 & 0.139 & 0.210 & 1.1 & 0.374 & 0.072 & 0.414\tabularnewline
W3C VCDM & jsonld (as specified) & verify & 0.217 & 0.035 & 0.209 & 0.241 & 1.7 & 0.568 & 0.120 & 0.471\tabularnewline
W3C VCDM (JCS) & canonicalize & sign & 0.048 & 0.021 & 0.044 & 0.061 & 0.5 & 0.127 & 0.033 & 0.122\tabularnewline
W3C VCDM (JCS) & canonicalize & verify & 0.125 & 0.018 & 0.122 & 0.137 & 1.9 & 0.362 & 0.086 & 0.277\tabularnewline
mdoc & cbor-x & sign & 0.127 & 0.081 & 0.112 & 0.155 & 0.8 & 0.282 & 0.059 & 0.367\tabularnewline
mdoc & cbor-x & verify & 0.140 & 0.046 & 0.131 & 0.161 & 1.0 & 0.367 & 0.081 & 0.446\tabularnewline
\midrule
\multicolumn{11}{@{}l}{\textit{Supplementary: reference implementation and canonicalization-free lower bounds}}\tabularnewline
SD-JWT VC & jose (reference) & sign & 0.112 & 0.068 & 0.101 & 0.137 & 2.1 & 0.211 & 0.053 & 0.232\tabularnewline
SD-JWT VC & jose (reference) & verify & 0.184 & 0.079 & 0.174 & 0.200 & 2.0 & 0.519 & 0.106 & 0.367\tabularnewline
W3C VCDM & canonicalization removed & sign & 0.043 & 0.025 & 0.041 & 0.054 & 0.5 & 0.122 & 0.032 & 0.099\tabularnewline
W3C VCDM & canonicalization removed & verify & 0.122 & 0.025 & 0.119 & 0.132 & 1.6 & 0.354 & 0.085 & 0.230\tabularnewline
mdoc & hand-written CBOR & sign & 0.125 & 0.071 & 0.111 & 0.151 & 1.9 & 0.293 & 0.058 & 0.335\tabularnewline
mdoc & hand-written CBOR & verify (COSE only) & 0.094 & 0.014 & 0.091 & 0.108 & 0.5 & 0.273 & 0.065 & 0.277\tabularnewline
\midrule
\multicolumn{11}{@{}l}{\textit{Canonicalization measured on its own}}\tabularnewline
W3C VCDM & one pass (document) & --- & 0.036 & 0.015 & 0.033 & 0.049 & 4.4 & 0.087 & 0.016 & 0.114\tabularnewline
W3C VCDM & both passes (as specified) & --- & 0.070 & 0.028 & 0.063 & 0.094 & 1.8 & 0.166 & 0.032 & 0.221\tabularnewline
\bottomrule
\end{tabular}
\end{table*}

\begin{table}[!t]
\centering
\caption{Cryptographic primitive baseline: the signature operations alone, over a fixed 256-byte message, with no credential structure (p50 in ms, medians across N=2,000 $\times$ 5 runs), in all four environments. Which of Ed25519 and P-256 verifies faster is a property of the runtime and the machine rather than of the algorithm: Node.js verifies P-256 faster than Ed25519 in A, B and C but not in D, where the order reverses, while Go verifies Ed25519 faster in all four.}\label{tab:prim}
\footnotesize\setlength{\tabcolsep}{5pt}
\begin{tabular}{@{}l c r r r@{}}
\toprule
& & \multicolumn{3}{c}{\textbf{p50 (ms)}}\tabularnewline
\cmidrule(lr){3-5}
\textbf{Primitive} & \textbf{Env.} & \textbf{Node} & \textbf{Go} & \textbf{Python}\tabularnewline
\midrule
Ed25519 sign & A & 0.038 & 0.028 & 0.036\tabularnewline
 & B & 0.110 & 0.092 & 0.105\tabularnewline
 & C & 0.029 & 0.022 & 0.029\tabularnewline
 & D & 0.088 & 0.063 & 0.077\tabularnewline
\addlinespace
Ed25519 verify & A & 0.116 & 0.065 & 0.116\tabularnewline
 & B & 0.343 & 0.224 & 0.344\tabularnewline
 & C & 0.084 & 0.051 & 0.082\tabularnewline
 & D & 0.243 & 0.142 & 0.240\tabularnewline
\addlinespace
ECDSA P-256 sign & A & 0.046 & 0.026 & 0.028\tabularnewline
 & B & 0.123 & 0.073 & 0.079\tabularnewline
 & C & 0.029 & 0.021 & 0.019\tabularnewline
 & D & 0.163 & 0.079 & 0.099\tabularnewline
\addlinespace
ECDSA P-256 verify & A & 0.091 & 0.079 & 0.073\tabularnewline
 & B & 0.271 & 0.237 & 0.228\tabularnewline
 & C & 0.064 & 0.057 & 0.054\tabularnewline
 & D & 0.317 & 0.250 & 0.252\tabularnewline
\addlinespace
SHA-256 & A & 0.002 & 0.000 & 0.001\tabularnewline
 & B & 0.005 & 0.003 & 0.001\tabularnewline
 & C & 0.001 & 0.000 & 0.000\tabularnewline
 & D & 0.009 & 0.003 & 0.003\tabularnewline
\bottomrule
\end{tabular}
\end{table}

At p50 in Environment~A, signing costs 0.038\,ms for SD-JWT VC, 0.112\,ms for mdoc and 0.139\,ms for W3C VCDM; verification costs 0.117\,ms for SD-JWT VC, 0.131\,ms for mdoc and 0.209\,ms for W3C VCDM. Of these relations, the one that holds in all four environments is W3C VCDM's distance from the other two: its verification is 1.40--1.90$\times$ SD-JWT VC's across A, B, C and D, and the slowest of the three in every environment. The mdoc-to-SD-JWT VC verification ratio, by contrast, is 1.12 in A, 1.05 in B, 0.95 in C and 1.80 in D---indistinguishable in B and C, 12\% apart in A, and 80\% apart in D. Two properties of the measured scope bear on how these figures should be read: the W3C VCDM rows include the canonicalization of the proof options that \texttt{eddsa-rdfc-2022} requires, and mdoc verification includes CBOR decoding and per-element digest matching rather than the COSE signature check alone, so each format is measured at the scope its specification defines.

Table~\ref{tab:prim} is what allows the rest of this section to be stated rather than conjectured, and having it in four environments is what turns the mdoc--SD-JWT VC spread above into an explanation. Ed25519 verification costs 0.116\,ms on its own in Node.js in Environment~A, and SD-JWT VC verification of a whole credential costs 0.117\,ms: the signature primitive is essentially all of it. Environment~D is where the primitives rank differently: there Node.js verifies Ed25519 in 0.243\,ms and P-256 in 0.317\,ms, the reverse of A, B and C, and signs P-256 1.86$\times$ slower than Ed25519 against 1.22$\times$ in Environment~A. Since mdoc is the only format using P-256 by default, that single primitive inversion accounts for the whole of mdoc's reversal in Environment~D, at both the micro level and the end-to-end level (Section~\ref{sec:e2e}). The W3C VCDM row with canonicalization removed costs 0.119\,ms, which is the same 0.116\,ms primitive plus the same small remainder. This is the correct reading of the \emph{canonicalization removed} configuration. That configuration feeds statically expanded N-Quads to the signature, which is not a faster canonicalizer but no canonicalizer at all---a verifier that skipped canonicalization could not establish that the signature covers the document it received. The figure is therefore a lower bound on what the pipeline costs once canonicalization is removed, and its being within a few percent of SD-JWT VC says only that both are then dominated by the same Ed25519 operation. We report it as a lower bound and draw no conclusion from it about the library.

The question whether the cost is the library's or the specification's is answerable, but from the cross-language data rather than from that row. Three RDFC-1.0 implementations were measured here: \texttt{jsonld} on Node.js, \texttt{json-gold} on Go and PyLD on Python. In Environment~A verification costs 0.209, 0.215 and 0.340\,ms respectively, and in each of the three languages, in each of the four environments---twelve combinations---W3C VCDM is the slowest of the three formats by a clear margin. The language ordering itself is not stable (Node.js is fastest in A, C and D, Go in B), which makes the invariance of the format ordering the more informative half of that result. The agreement among the three implementations is nonetheless weaker evidence than its arithmetic suggests, because all three descend from jsonld.js: they are ports of one design rather than independent implementations of the specification, so a shared implementation choice would show up in all three alike. Our replication adds two implementations that do not share that lineage---the Rust \texttt{rdf-canon} and the Java titanium-rdfc---and finds canonicalization costly in those as well; those runs are outside the measurement reported here, so we cite them as corroboration rather than as data of ours. The argument we actually rest on is the one in the next paragraph, which varies the canonicalization rather than the implementation. The comparison with JCS points the same way, and does so in every environment: \texttt{eddsa-jcs-2022} canonicalizes the same credential lexically rather than as an RDF dataset and verifies within 5\% of SD-JWT VC in Environments~A, B and C (within 0.1\% in C) and within 12\% in D (Table~\ref{tab:micro}; D is just outside our 10\% band, where the RDFC-1.0 configuration is 1.90$\times$ SD-JWT VC), so it is the choice of RDF canonicalization specifically, not JSON-LD and not Data Integrity as such, that carries the cost.

\begin{figure*}[!t]
\centering
\subfloat[Node.js]{\includegraphics[width=0.32\textwidth]{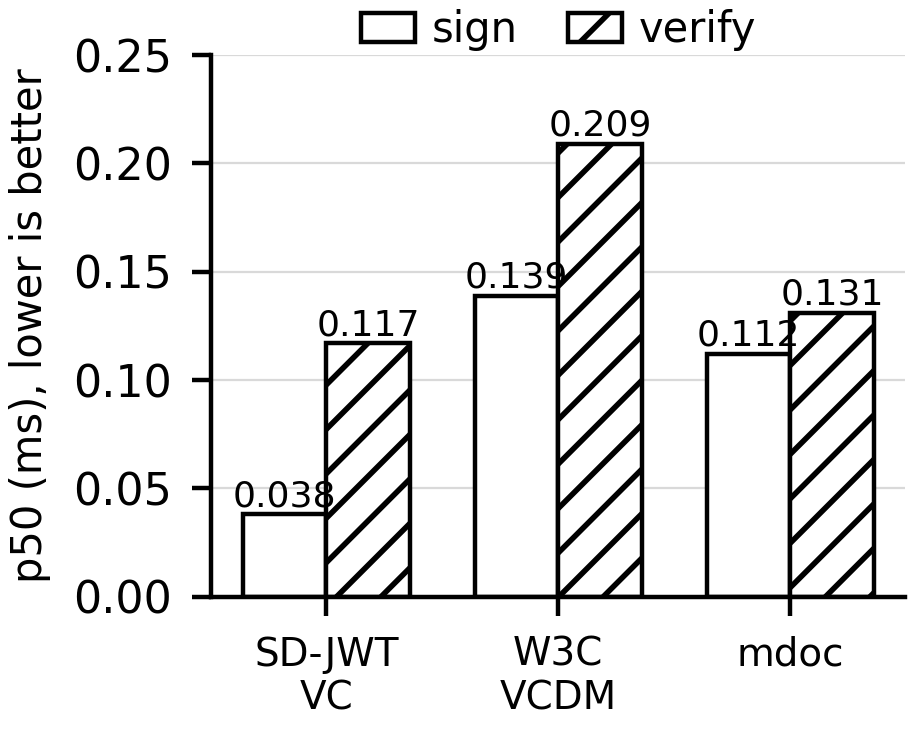}\label{fig:f1}}
\hfil
\subfloat[Python]{\includegraphics[width=0.32\textwidth]{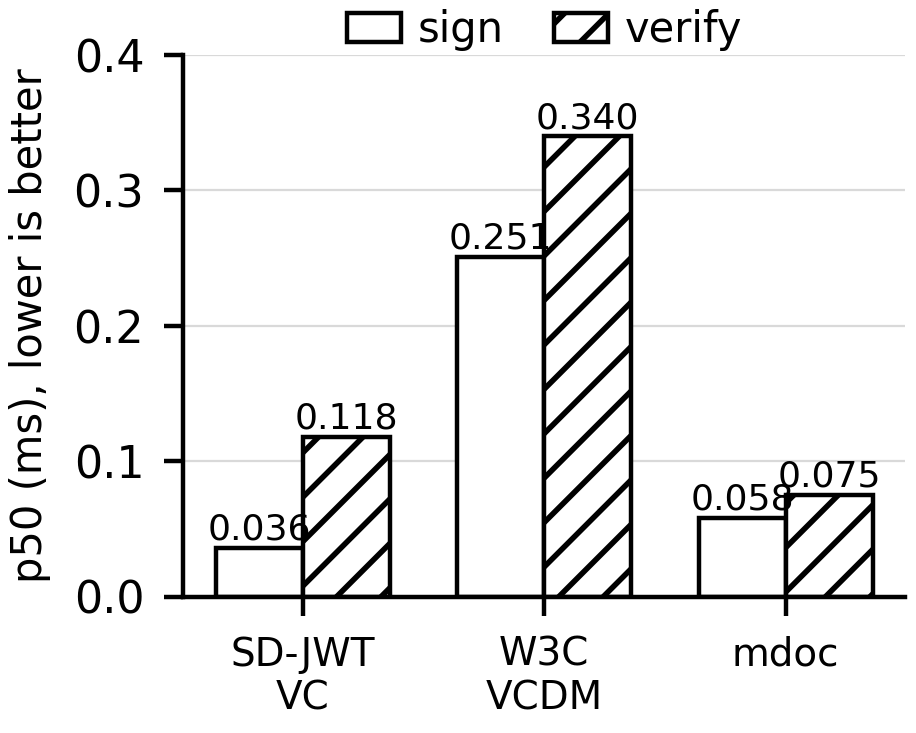}\label{fig:f2}}
\hfil
\subfloat[Go]{\includegraphics[width=0.32\textwidth]{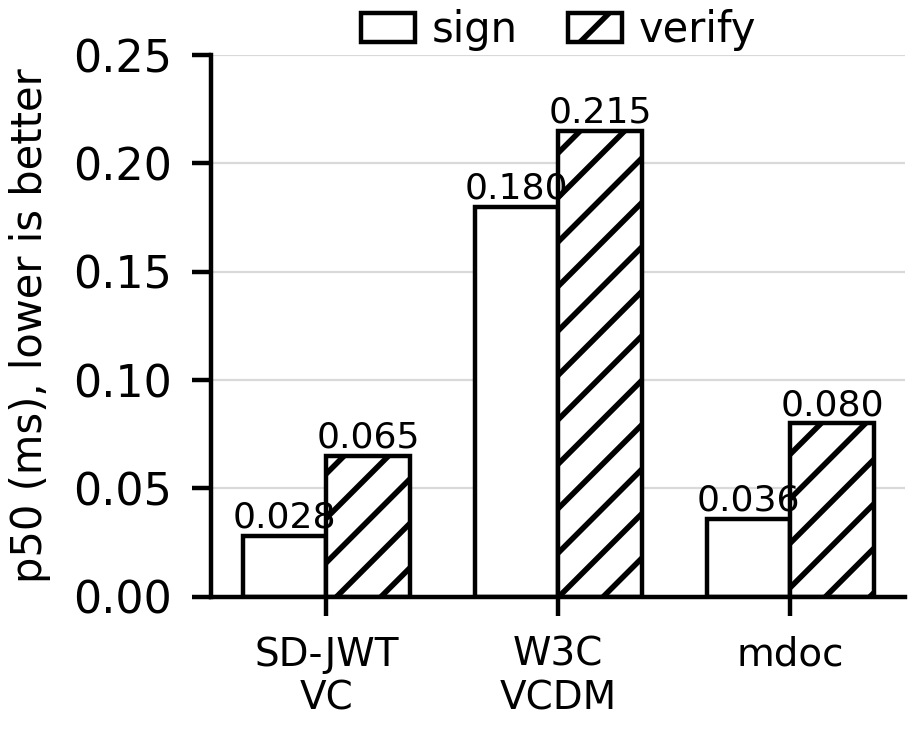}\label{fig:f3}}
\caption{Signing and verification of a single credential in Environment~A, in the three language runtimes: Node.js, Python (PyLD, cbor2, \texttt{cryptography}) and Go (json-gold, fxamacker/cbor, standard library). Bars are p50 in milliseconds (lower is better); hatched bars are verification. The vertical axis differs between panels. Data for (a) as in Table~\ref{tab:micro}. Note the reversal in (c): SD-JWT VC verifies faster than mdoc, which Table~\ref{tab:prim} attributes to Go's Ed25519 implementation.}\label{fig:lang}
\end{figure*}

The runtimes also decide which of SD-JWT VC and mdoc verifies faster, and Table~\ref{tab:prim} explains why. In Environment~A: in Go (Fig.~\ref{fig:f3}) SD-JWT VC verifies in 0.065\,ms against mdoc's 0.080\,ms; in Python (Fig.~\ref{fig:f2}) the order reverses, mdoc 0.075\,ms against SD-JWT VC 0.118\,ms; in Node.js (Fig.~\ref{fig:f1}) they are close, 0.117 against 0.131\,ms. Environment~D rearranges this pattern as well---Python there puts SD-JWT VC ahead (0.246 against 0.262\,ms), where in A, B and C it puts mdoc ahead---so the pair's ordering depends on the runtime and the machine jointly, and on neither alone. Two independent effects produce this. First, the signature primitives rank differently per runtime: Go verifies Ed25519 in 0.065\,ms and P-256 in 0.079\,ms, while Node.js and Python verify P-256 faster than Ed25519 (0.091 against 0.116\,ms, and 0.073 against 0.116\,ms). Go's standard-library Ed25519 is simply quicker than its P-256, which alone would favor SD-JWT VC there. Second, the non-cryptographic remainder differs by an order of magnitude: mdoc verification above its P-256 primitive costs 0.040\,ms in Node.js but 0.002\,ms in Python, because \texttt{cbor2} decodes in C while \texttt{cbor-x} does more work in JavaScript. The ordering of these two formats is therefore a property of the runtime's cryptographic and CBOR implementations, and any claim that one of them is faster needs the runtime attached to it.

W3C VCDM is slowest in all three languages in both operations, in all four environments, and the gap is not attributable to the runtime in the same way. Verification in Environment~A: Node.js 0.209\,ms, Go 0.215\,ms, Python 0.340\,ms, against 0.065--0.131\,ms for the other two formats. In Go, where the Ed25519 primitive is the fastest of any runtime measured, W3C VCDM is still 3.3$\times$ SD-JWT VC in Environment~A---and 2.5$\times$, 3.0$\times$ and 4.4$\times$ in B, C and D. The canonicalization requirement is what remains when the runtime and the machine are both varied.

\subsection{Serialization Speed (Without Cryptographic Operations)}\label{sec:serial}

Table~\ref{tab:t10} shows the pure serialization (encoding/decoding/canonicalization) speed without signature verification. This evaluates data transformation overhead excluding cryptographic processing. SD-JWT VC is evaluated for Base64URL encoding/decoding, W3C VCDM for JSON serialization/parsing and RDFC-1.0 canonicalization, and mdoc for CBOR encoding/decoding.

\begin{table*}[!t]
\centering
\caption{Serialization and canonicalization without any cryptographic operation, in all four environments (p50 in ms, medians across N=2,000 $\times$ 5 runs). The \emph{kind of work} column separates two things the table would otherwise conflate: encoding a payload, and producing the exact byte string the signature is computed over. The last column gives the size of the artifact produced, which differs between formats and is why absolute times across rows are not like for like.}\label{tab:t10}
\footnotesize\setlength{\tabcolsep}{3pt}
\begin{tabular}{@{}l l l r r r r r r l@{}}
\toprule
& & & \multicolumn{3}{c}{\textbf{Env.~A (ms)}} & \multicolumn{3}{c}{\textbf{p50 (ms)}} & \tabularnewline
\cmidrule(lr){4-6}\cmidrule(lr){7-9}
\textbf{Format} & \textbf{Operation} & \textbf{Kind of work} & \textbf{mean} & \textbf{$\sigma$} & \textbf{p50} & \textbf{B} & \textbf{C} & \textbf{D} & \textbf{Bytes (what)}\tabularnewline
\midrule
SD-JWT VC & Base64URL encode & encoding only & 0.003 & 0.014 & 0.002 & 0.004 & 0.001 & 0.005 & 322 (JWT, no Disclosures)\tabularnewline
SD-JWT VC & Base64URL decode & encoding only & 0.003 & 0.004 & 0.002 & 0.008 & 0.001 & 0.008 & 322 (JWT, no Disclosures)\tabularnewline
W3C VCDM & JSON encode & encoding only & 0.002 & 0.012 & 0.001 & 0.003 & 0.001 & 0.004 & ---\tabularnewline
W3C VCDM & JSON decode & encoding only & 0.002 & 0.002 & 0.002 & 0.004 & 0.001 & 0.005 & ---\tabularnewline
W3C VCDM & RDFC-1.0 (\texttt{jsonld}) & signing input & 0.055 & 0.097 & 0.042 & 0.120 & 0.021 & 0.156 & 464 (N-Quads, signing input)\tabularnewline
W3C VCDM (JCS) & JCS (RFC 8785) & signing input & 0.007 & 0.049 & 0.006 & 0.012 & 0.003 & 0.014 & 612 (canonical JSON)\tabularnewline
mdoc & CBOR encode & encoding only & 0.004 & 0.034 & 0.002 & 0.007 & 0.002 & 0.006 & 497 (CBOR)\tabularnewline
mdoc & CBOR decode & encoding only & 0.004 & 0.021 & 0.003 & 0.008 & 0.002 & 0.008 & 497 (CBOR)\tabularnewline
\bottomrule
\end{tabular}
\end{table*}

In encoding, SD-JWT VC (Base64URL) achieved p50: 0.002\,ms, W3C VCDM (JSON.stringify) 0.001\,ms, and mdoc (CBOR via cbor-x) 0.002\,ms---all on the order of microseconds. The last column (Bytes) reports the JWT token length for SD-JWT VC, the RDFC-1.0-normalized N-Quads for W3C VCDM, the JCS-canonicalized JSON for W3C VCDM (JCS), and the CBOR-encoded bytes for mdoc.

In canonicalization processing, W3C VCDM's RDFC-1.0 canonicalization (jsonld library) took p50: 0.042\,ms in Environment~A while JCS (JSON Canonicalization Scheme, RFC 8785) took 0.006\,ms, approximately 7$\times$ faster; the same ratio is 10.1$\times$, 6.2$\times$ and 10.9$\times$ in B, C and D, so RDFC-1.0 is the more expensive of the two ways of producing a signing input in every environment, by between six and eleven times. RDFC-1.0 performs RDF graph canonicalization including blank node relabeling, making it computationally more expensive than JCS's key sorting and whitespace removal.

Encoding and decoding cost 0.006\,ms or less in every format in Environment~A and 0.014\,ms or less in all four, one to two orders of magnitude below signing and verification (Table~\ref{tab:micro}: p50 0.117--0.209\,ms for verification in Environment~A). The choice between JSON and CBOR as a wire encoding is therefore not what separates the formats; the signature algorithm and the presence or absence of canonicalization are. The contrast within the table makes the point: RDFC-1.0 costs 0.042\,ms where CBOR encoding costs 0.002\,ms, and the two are not alternatives at the same layer---one produces a signing input, the other only a wire form. Note that this RDFC-1.0 measurement applies to a simple credential with a single blank node; for complex credentials, canonicalization becomes the dominant cost (Section~\ref{sec:complex}).

\subsection{Attribute Count Scaling Evaluation}\label{sec:scaling}

\begin{figure}[!t]
\centering
\includegraphics[width=\columnwidth]{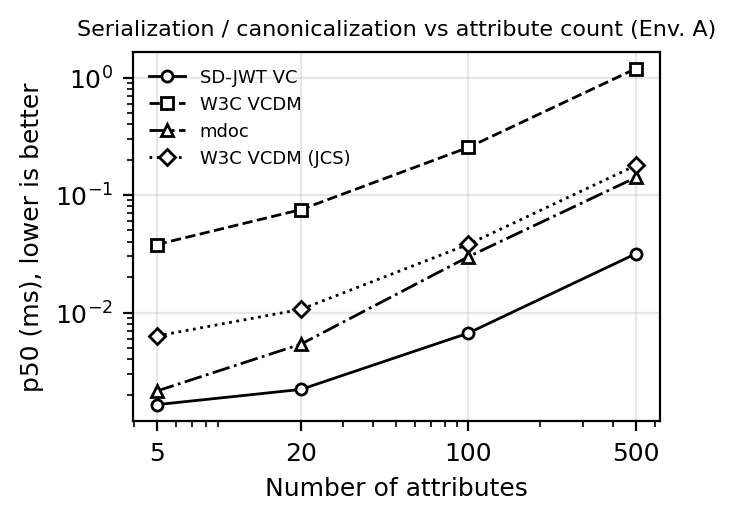}
\caption{Attribute-count scaling, Environment~A. Both axes logarithmic: horizontal, number of attributes (5, 20, 100, 500); vertical, p50 in milliseconds, lower is better. Series are distinguished by both line style and marker. The W3C VCDM series is RDFC-1.0 canonicalization; the others are encoding or JCS canonicalization (Table~\ref{tab:t11}).}\label{fig:f4}
\end{figure}

Table~\ref{tab:t11} and Fig.~\ref{fig:f4} show the change in serialization speed (without cryptographic processing; W3C VCDM includes RDFC-1.0 canonicalization via the jsonld library) when varying the number of attributes to 5, 20, 100, and 500. Measured with process.hrtime.bigint() at N=2,000 (N=400 for W3C VCDM at 100+ attributes) $\times$ 5 runs.

\begin{table*}[!t]
\centering
\caption{Attribute-count scaling of the serialization or canonicalization step, in all four environments (medians across N=2,000 $\times$ 5 runs; p50 in ms, ops/s from p50). For W3C VCDM the measured step is RDFC-1.0 canonicalization, for the others it is encoding, so the columns are comparable within a format across attribute counts and not between formats. $\dagger$ marks conditions whose iteration count was reduced to fit the time budget; the result files record the count actually completed.}\label{tab:t11}
\footnotesize\setlength{\tabcolsep}{3pt}
\begin{tabular}{@{}l r r r r r r r r r r@{}}
\toprule
& & \multicolumn{5}{c}{\textbf{Env.~A (ms except ops/s)}} & \multicolumn{3}{c}{\textbf{p50 (ms)}} & \tabularnewline
\cmidrule(lr){3-7}\cmidrule(lr){8-10}
\textbf{Format} & \textbf{Attrs} & \textbf{mean} & \textbf{ops/s} & \textbf{$\sigma$} & \textbf{p50} & \textbf{p95} & \textbf{B} & \textbf{C} & \textbf{D} & \textbf{Size (bytes)}\tabularnewline
\midrule
SD-JWT VC & 5 & 0.003 & 608,643 & 0.014 & 0.002 & 0.004 & 0.004 & 0.001 & 0.005 & 272\tabularnewline
W3C VCDM & 5 & 0.053 & 26,413 & 0.133 & 0.038 & 0.073 & 0.112 & 0.020 & 0.142 & 681\tabularnewline
W3C VCDM (JCS) & 5 & 0.007 & 158,428 & 0.008 & 0.006 & 0.008 & 0.014 & 0.004 & 0.017 & 581\tabularnewline
mdoc & 5 & 0.003 & 466,418 & 0.015 & 0.002 & 0.004 & 0.005 & 0.001 & 0.005 & 355\tabularnewline
\addlinespace
SD-JWT VC & 20 & 0.003 & 451,671 & 0.004 & 0.002 & 0.004 & 0.005 & 0.001 & 0.006 & 732\tabularnewline
W3C VCDM & 20 & 0.084 & 13,304 & 0.058 & 0.075 & 0.104 & 0.196 & 0.041 & 0.241 & 1,716\tabularnewline
W3C VCDM (JCS) & 20 & 0.011 & 93,985 & 0.006 & 0.011 & 0.011 & 0.023 & 0.006 & 0.028 & 926\tabularnewline
mdoc & 20 & 0.006 & 185,874 & 0.002 & 0.005 & 0.008 & 0.014 & 0.004 & 0.015 & 1,300\tabularnewline
\addlinespace
SD-JWT VC & 100 & 0.007 & 149,633 & 0.007 & 0.007 & 0.010 & 0.016 & 0.004 & 0.020 & 3,185\tabularnewline
W3C VCDM $\dagger$ & 100 & 0.278 & 3,930 & 0.061 & 0.254 & 0.427 & 0.645 & 0.143 & 0.744 & 7,236\tabularnewline
W3C VCDM (JCS) & 100 & 0.039 & 26,281 & 0.007 & 0.038 & 0.041 & 0.074 & 0.024 & 0.078 & 2,766\tabularnewline
mdoc & 100 & 0.030 & 33,506 & 0.005 & 0.030 & 0.033 & 0.075 & 0.020 & 0.070 & 6,417\tabularnewline
\addlinespace
SD-JWT VC & 500 & 0.035 & 31,617 & 0.031 & 0.032 & 0.046 & 0.082 & 0.022 & 0.083 & 15,452\tabularnewline
W3C VCDM $\dagger$ & 500 & 1.245 & 840 & 0.151 & 1.191 & 1.446 & 2.986 & 0.663 & 2.844 & 34,836\tabularnewline
W3C VCDM (JCS) & 500 & 0.182 & 5,598 & 0.012 & 0.179 & 0.191 & 0.337 & 0.117 & 0.343 & 11,966\tabularnewline
mdoc & 500 & 0.138 & 7,091 & 0.028 & 0.141 & 0.155 & 0.356 & 0.096 & 0.329 & 32,262\tabularnewline
\bottomrule
\end{tabular}
\end{table*}

RDFC-1.0 canonicalization of a W3C VCDM credential rose from p50: 0.038\,ms at 5 attributes to 1.191\,ms at 500, a roughly 30-fold increase for a 100-fold increase in attributes. The growth rate itself rises with scale: in Environment~A the per-interval exponent (the slope on log-log axes) increases monotonically from 0.49 for 5--20 attributes to 0.76 for 20--100 and 0.96 for 100--500, so the cost approaches linear growth at the top of the range. All four environments reproduce that shape with the exponent rising monotonically over the same three intervals---0.40/0.74/0.95 in B, 0.52/0.78/0.95 in C and 0.38/0.70/0.83 in D---and the end-to-end multiple from 5 to 500 attributes is 32$\times$, 27$\times$, 34$\times$ and 20$\times$. Sub-linear growth that approaches linearity is therefore the robust statement; the exponent itself is environment-specific to within about 0.1. Over the same range SD-JWT VC's Base64URL encoding rose from 0.002 to 0.032\,ms, JCS canonicalization from 0.006 to 0.179\,ms and CBOR encoding from 0.002 to 0.141\,ms. The ratio of W3C VCDM's 1.191\,ms to SD-JWT VC's 0.032\,ms at 500 attributes (about 37$\times$) compares a canonicalization against an encoding and should be read as such; the comparison between like steps is W3C VCDM's 1.191\,ms against JCS's 0.179\,ms in Environment~A, a factor of 6.7 between two ways of producing a signing input for the same credential, and 8.9, 5.7 and 8.3 in B, C and D. RDFC-1.0 canonicalization involves blank node relabeling and RDF dataset processing, so its cost grows faster than the lexical canonicalization of JCS or the binary encoding of mdoc as the number of statements (attributes) increases.

Regarding payload size as well, W3C VCDM (RDFC-1.0-normalized N-Quads) was 34,836 bytes at 500 attributes, approximately 2.3$\times$ that of SD-JWT VC (15,452 bytes). This is because the RDF data model represents each attribute as a separate triple (subject-predicate-object), resulting in larger serialized output. mdoc was 32,262 bytes at 500 attributes, comparable to W3C VCDM, at the cost of human readability since it is binary CBOR.

\subsection{Selective Disclosure Performance}\label{sec:seldisc}

\begin{figure}[!t]
\centering
\includegraphics[width=\columnwidth]{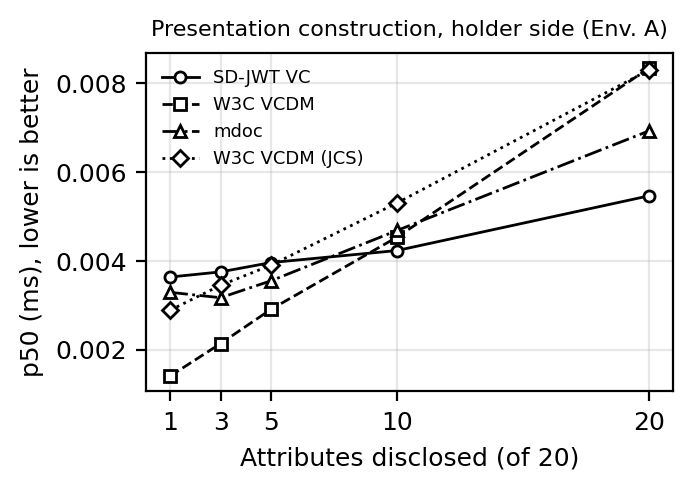}
\caption{Cost of subset selection, holder side only, Environment~A. Horizontal axis: attributes disclosed out of 20; vertical axis: p50 in milliseconds, lower is better. This is not the complete \emph{present} step of Table~\ref{tab:e2e}: the W3C VCDM series excludes re-canonicalization and re-signing (Table~\ref{tab:t12}).}\label{fig:f5}
\end{figure}

Table~\ref{tab:t12} and Fig.~\ref{fig:f5} show the cost of \emph{subset selection}---choosing which attributes to reveal and re-encoding that subset---for 1, 3, 5, 10 and 20 disclosed attributes out of a total of 20. Two names are worth keeping apart here. The \emph{present} column of Table~\ref{tab:e2e} is the complete holder-side presentation step, which for \texttt{eddsa-rdfc-2022} includes re-canonicalizing and re-signing the subset; \emph{subset selection}, measured in this subsection, is only the first part of it, and for SD-JWT VC and mdoc is nearly all of it. The two are not the same measurement and their figures should not be compared across the tables. Measurements use process.hrtime.bigint() (nanosecond resolution) at N=2,000 $\times$ 5 runs; subset selection is a microsecond-scale operation, so resolving differences at this level requires a timer with sub-microsecond resolution.

\begin{table*}[!t]
\centering
\caption{Cost of \emph{subset selection}, by number of disclosed attributes, in all four environments (p50 in ms, medians across N=2,000 $\times$ 5 runs). This is the holder side only, and only the selection part of it: the complete presentation step is the \emph{present} column of Table~\ref{tab:e2e}, which for W3C VCDM additionally re-canonicalizes and re-signs the subset (0.134\,ms there in Environment~A) and which is the figure to quote for presentation cost. The verifier-side cost of matching what was disclosed is also in Table~\ref{tab:e2e}; issuer-signature-preserving derivation is in Table~\ref{tab:sd}.}\label{tab:t12}
\footnotesize\setlength{\tabcolsep}{4pt}
\begin{tabular}{@{}l c r r r r r@{}}
\toprule
& & \multicolumn{5}{c}{\textbf{p50 (ms)}}\tabularnewline
\cmidrule(lr){3-7}
\textbf{Format} & \textbf{Env.} & \textbf{1 of 20} & \textbf{3 of 20} & \textbf{5 of 20} & \textbf{10 of 20} & \textbf{20 of 20}\tabularnewline
\midrule
SD-JWT VC & A & 0.004 & 0.004 & 0.004 & 0.004 & 0.005\tabularnewline
 & B & 0.010 & 0.010 & 0.011 & 0.011 & 0.014\tabularnewline
 & C & 0.002 & 0.002 & 0.002 & 0.002 & 0.003\tabularnewline
 & D & 0.011 & 0.011 & 0.012 & 0.013 & 0.016\tabularnewline
\addlinespace
W3C VCDM (subset selection only) & A & 0.001 & 0.002 & 0.003 & 0.005 & 0.008\tabularnewline
 & B & 0.004 & 0.005 & 0.007 & 0.010 & 0.018\tabularnewline
 & C & 0.001 & 0.001 & 0.002 & 0.003 & 0.005\tabularnewline
 & D & 0.004 & 0.006 & 0.007 & 0.011 & 0.019\tabularnewline
\addlinespace
W3C VCDM (JCS) & A & 0.003 & 0.003 & 0.004 & 0.005 & 0.008\tabularnewline
 & B & 0.006 & 0.008 & 0.009 & 0.012 & 0.017\tabularnewline
 & C & 0.002 & 0.002 & 0.002 & 0.003 & 0.005\tabularnewline
 & D & 0.008 & 0.009 & 0.010 & 0.014 & 0.021\tabularnewline
\addlinespace
mdoc & A & 0.003 & 0.003 & 0.004 & 0.005 & 0.007\tabularnewline
 & B & 0.009 & 0.008 & 0.009 & 0.012 & 0.018\tabularnewline
 & C & 0.002 & 0.002 & 0.002 & 0.003 & 0.005\tabularnewline
 & D & 0.010 & 0.009 & 0.011 & 0.014 & 0.021\tabularnewline
\bottomrule
\end{tabular}
\end{table*}

Subset selection takes 1--8 microseconds at p50 for every format in Environment~A, and 1--21 microseconds across all four, one to two orders of magnitude below verification in every environment (Table~\ref{tab:e2e}: p50 0.081--0.583\,ms). W3C VCDM, its JCS variant and mdoc grow monotonically with the number of attributes revealed, from about 0.001--0.003\,ms at one to 0.007--0.008\,ms at twenty, while SD-JWT VC is nearly flat at 0.004--0.005\,ms because presentation only concatenates the selected Disclosures. Two qualifications keep this table from being read for more than it shows. First, it measures the holder's side only; the verifier's cost of matching what was disclosed is in Table~\ref{tab:e2e}, and for SD-JWT VC and mdoc that is where the disclosure cost actually lands. Second, the W3C VCDM row measures selecting and re-encoding the subset, not re-canonicalizing and re-signing it, which is what an \texttt{eddsa-rdfc-2022} holder must actually do and which Table~\ref{tab:e2e} measures at 0.134\,ms---two orders of magnitude more than this row. Subset selection is not a dominant cost for any format, but for W3C VCDM that statement does not extend to the complete \emph{present} step.

\subsection{Ed25519 Unified Algorithm Benchmark}\label{sec:unified-micro}

The default configuration of each ecosystem uses Ed25519 for SD-JWT VC and W3C VCDM and ECDSA P-256 for mdoc, so signature algorithm differences constitute a confounding factor in the step-level numbers. Table~\ref{tab:t13} shows the results of unifying all three formats under Ed25519, with the payload also unified to 5 attributes, to isolate this confound. The libraries are those of Table~\ref{tab:micro}, at the narrower scope stated in the caption (only mdoc's COSE alg changes to $-8$, EdDSA), measured with process.hrtime.bigint() at N=2,000 $\times$ 5 runs. Note that the mdoc Ed25519 values represent a configuration ISO/IEC 18013-5 permits but deployed mDL profiles do not use by default (ES256 is the usual choice).

\begin{table*}[!t]
\centering
\caption{Signing and verification of a 5-attribute credential with Ed25519 for all three formats, in all four environments (p50 in ms, medians across N=2,000 $\times$ 5 runs; mdoc on COSE alg $-8$, a configuration ISO/IEC 18013-5 permits but deployed mDL profiles do not use by default). \textbf{The scope here is narrower than in Table~\ref{tab:micro}}: the mdoc rows verify the COSE\_Sign1 signature alone, without CBOR decoding or digest matching, and the W3C VCDM rows canonicalize once rather than twice. The figures are therefore comparable within this table but not against Table~\ref{tab:micro}; for an algorithm substitution measured at the full scope, see Table~\ref{tab:unified-e2e}.}\label{tab:t13}
\footnotesize\setlength{\tabcolsep}{4pt}
\begin{tabular}{@{}l l r r r r r r r@{}}
\toprule
& & \multicolumn{4}{c}{\textbf{Env.~A (ms)}} & \multicolumn{3}{c}{\textbf{p50 (ms)}}\tabularnewline
\cmidrule(lr){3-6}\cmidrule(lr){7-9}
\textbf{Format} & \textbf{Op} & \textbf{mean} & \textbf{$\sigma$} & \textbf{p50} & \textbf{p95} & \textbf{B} & \textbf{C} & \textbf{D}\tabularnewline
\midrule
SD-JWT VC & sign & 0.042 & 0.024 & 0.040 & 0.052 & 0.116 & 0.029 & 0.099\tabularnewline
SD-JWT VC & verify & 0.119 & 0.005 & 0.118 & 0.130 & 0.349 & 0.084 & 0.253\tabularnewline
W3C VCDM & sign & 0.120 & 0.044 & 0.107 & 0.154 & 0.286 & 0.057 & 0.295\tabularnewline
W3C VCDM & verify & 0.189 & 0.027 & 0.180 & 0.209 & 0.493 & 0.109 & 0.381\tabularnewline
mdoc & sign & 0.059 & 0.054 & 0.053 & 0.071 & 0.158 & 0.037 & 0.140\tabularnewline
mdoc & verify & 0.116 & 0.003 & 0.115 & 0.125 & 0.339 & 0.081 & 0.209\tabularnewline
\bottomrule
\end{tabular}
\end{table*}

Under Ed25519, SD-JWT VC remains the fastest signer (p50: 0.040\,ms), followed by mdoc (0.053\,ms) and W3C VCDM (0.107\,ms). In verification SD-JWT VC (0.118\,ms) and mdoc (0.115\,ms) differ by 2\%, well inside the 10\% band defined in Section~\ref{sec:env}, and we therefore treat them as indistinguishable here; W3C VCDM is 1.5$\times$ either of them at 0.180\,ms. Unifying the algorithm has the same effect in all four environments: the mdoc-to-SD-JWT VC verification ratio becomes 0.98, 0.97, 0.97 and 0.82 in A, B, C and D. The first three are inside the 10\% band; the 1.80$\times$ that separates them in Environment~D under their default algorithms (Section~\ref{sec:lang}) does not survive holding the primitive fixed, though the two do not become indistinguishable there either---mdoc is the faster of the pair by 18\%, in the opposite direction and at a fifth of the magnitude. W3C VCDM meanwhile remains 1.29--1.53$\times$ SD-JWT VC in all four. In other words, the mdoc--SD-JWT VC difference is a statement about primitives and machines, and the W3C VCDM difference is a statement about the pipeline. The mdoc figure of 0.115\,ms should not be read against the 0.131\,ms of Table~\ref{tab:micro}: this table's mdoc rows verify the COSE signature alone, and 0.115\,ms is within noise of the Ed25519 primitive itself (0.116\,ms, Table~\ref{tab:prim}).

Measured at equal scope, the substitution goes the other way. In the end-to-end comparison of Table~\ref{tab:unified-e2e}, mdoc verification in Environment~A costs 0.140\,ms under Ed25519 against 0.127\,ms under P-256, which tracks the primitives (0.116 against 0.091\,ms). The substitution follows the primitives in each environment, including where they invert: in B it costs mdoc (0.414 against 0.357\,ms) and in C too (0.096 against 0.081\,ms), while in D, where Node.js verifies Ed25519 faster than P-256, it helps mdoc (0.308 against 0.376\,ms). The decoding and digest matching that \texttt{cbor-x} adds on top of the primitive is positive under either algorithm in every environment, while a COSE-only check stays within a few percent of the primitive. That is, the verification speed advantage of mdoc's default configuration derives partly from OpenSSL's fast ECDSA P-256 verification rather than from the CBOR/COSE structure. Meanwhile, W3C VCDM's residual gap (1.5$\times$ SD-JWT VC under a unified algorithm) stems from the RDFC-1.0 canonicalization and JSON-LD processing inherent in its verification pipeline, and does not disappear when the algorithm is unified.

\subsection{RDFC-1.0 Canonicalization on Complex Credentials}\label{sec:complex}

Table~\ref{tab:t14} contrasts the simple credential used in the main benchmark (5 quads, one blank node) with complex credentials based on production schemas. The evaluated credentials are: a 1EdTech Open Badges v3.0 (OB3) AchievementCredential, widely used in the education domain (containing id-less nodes such as criteria, alignment, and result); a sample modeled on the OB3-profile academic degree credentials issued by the Digital Credentials Commons (DCC); and synthetic credentials with 10 and 50 id-less child nodes to isolate the effect of blank node count. All JSON-LD contexts are statically embedded; no network I/O is included in the measurements.

\begin{table*}[!t]
\centering
\caption{RDFC-1.0 canonicalization of complex credentials in all four environments (\texttt{jsonld} 8.3.3 over \texttt{rdf-canonize} 3.4.0, p50 in ms, medians across 5 runs; \emph{vs simple} is the ratio to that environment's own simple-credential baseline). $\dagger$ marks conditions whose iteration count was reduced to fit the time budget. The ordering of the credentials is identical in all four environments; the multiple is not, and the Open Badges ratio ranges from 31.1$\times$ to 49.5$\times$ across them, which is why we quote it as a range rather than as a constant. Bold marks the Open Badges multiples, the figures quoted in the text and the abstract.}\label{tab:t14}
\footnotesize\setlength{\tabcolsep}{3pt}
\begin{tabular}{@{}l r r r r r r r r r r@{}}
\toprule
& & & \multicolumn{2}{c}{\textbf{Env.~A}} & \multicolumn{2}{c}{\textbf{Env.~B}} & \multicolumn{2}{c}{\textbf{Env.~C}} & \multicolumn{2}{c}{\textbf{Env.~D}}\tabularnewline
\cmidrule(lr){4-5}\cmidrule(lr){6-7}\cmidrule(lr){8-9}\cmidrule(lr){10-11}
\textbf{Credential} & \textbf{Quads} & \textbf{BN} & \textbf{p50 (ms)} & \textbf{vs simple} & \textbf{p50 (ms)} & \textbf{vs simple} & \textbf{p50 (ms)} & \textbf{vs simple} & \textbf{p50 (ms)} & \textbf{vs simple}\tabularnewline
\midrule
Simple credential & 5 & 1 & 0.047 & 1.0$\times$ & 0.122 & 1.0$\times$ & 0.021 & 1.0$\times$ & 0.170 & 1.0$\times$\tabularnewline
Open Badges v3.0 $\dagger$ & 32 & 4 & 1.603 & \textbf{34.1$\times$} & 3.924 & \textbf{32.1$\times$} & 1.039 & \textbf{49.5$\times$} & 5.298 & \textbf{31.1$\times$}\tabularnewline
DCC-style degree credential $\dagger$ & 21 & 2 & 1.092 & 23.2$\times$ & 2.650 & 21.7$\times$ & 0.709 & 33.8$\times$ & 2.772 & 16.3$\times$\tabularnewline
Synthetic (10 blank nodes) & 43 & 10 & 0.187 & 4.0$\times$ & 0.501 & 4.1$\times$ & 0.098 & 4.7$\times$ & 0.526 & 3.1$\times$\tabularnewline
Synthetic (50 blank nodes) $\dagger$ & 203 & 50 & 0.748 & 15.9$\times$ & 1.864 & 15.3$\times$ & 0.428 & 20.4$\times$ & 1.991 & 11.7$\times$\tabularnewline
\bottomrule
\end{tabular}
\end{table*}

\begin{figure}[!t]
\centering
\includegraphics[width=\columnwidth]{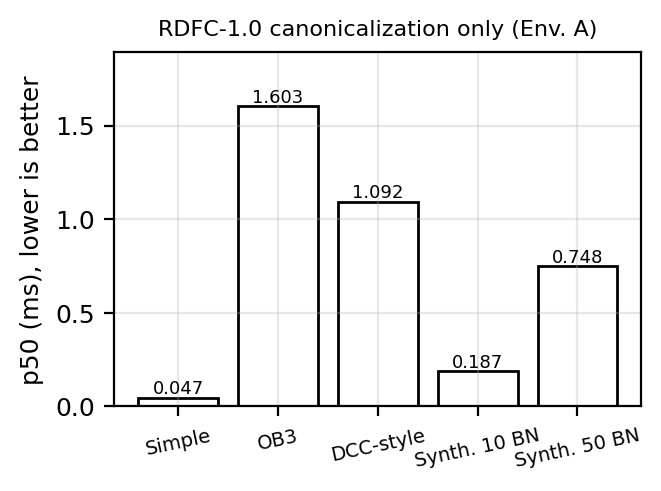}
\caption{RDFC-1.0 canonicalization of complex credentials, Environment~A. Vertical axis: p50 in milliseconds, lower is better. This is the canonicalization step alone, not a verification pipeline. Data as in Table~\ref{tab:t14}.}\label{fig:f6}
\end{figure}

Canonicalizing the OB3 credential took p50: 1.603\,ms in Environment~A, roughly 34$\times$ the simple credential (0.047\,ms); the DCC-style credential took 23$\times$ (1.092\,ms). Fig.~\ref{fig:f6} plots the five credentials on a common axis. Across the four environments the OB3 multiple is 34.1$\times$, 32.1$\times$, 49.5$\times$ and 31.1$\times$ and the DCC-style multiple 23.2$\times$, 21.7$\times$, 33.8$\times$ and 16.3$\times$; the ordering of the five credentials is identical in all four, so what the table supports is that production schemas cost one to two orders of magnitude more than the simple credential, with the exact multiple a property of the machine. The comparison that is worth making carefully is with the rest of the pipeline, because the two quantities are not the same kind of thing: 1.603\,ms is the cost of canonicalization alone, whereas the end-to-end verification step of Table~\ref{tab:e2e} (20 attributes, 5 disclosed) is 0.212\,ms for the whole W3C VCDM pipeline. Canonicalizing one production credential therefore costs about eight times that verification step in Environment~A, and 6.7, 8.3 and 10.9 times in B, C and D---a real and large effect in every environment, but it is a step-against-pipeline comparison and we label it as such rather than calling it an order of magnitude above ``the verification pipeline''.

Separating the two contributing factors matters more than their sum. The first is expansion of OB3's large JSON-LD context, with its hundreds of term definitions; the second is blank node canonical labeling. The synthetic credentials isolate the second: in Environment~A 10 id-less nodes cost 0.187\,ms and 50 cost 0.748\,ms, a factor of 4.0 for five times the nodes, an exponent of 0.86 over that interval; the same exponent is 0.82, 0.92 and 0.83 in B, C and D, so blank node cost grows sub-linearly in the number of nodes in all four. OB3 itself points the same way: it has only 4 blank nodes yet costs more than the synthetic credential with 50---2.1$\times$ more in Environment~A and 2.1--2.7$\times$ more in the other three. The two inputs differ in quad count as well (32 against 203), so this does not isolate expansion from labeling; separating them would need the expansion step timed on its own. What it does show, in every environment measured, is that blank node labeling is not the dominant term for these schemas. A mitigation expressed as an upper bound on blank node count would therefore miss the cost almost entirely; pre-compiled contexts and expansion caching are what applies. The multiples are largest in Environment~C, the fastest machine, whose baseline is 0.021\,ms, and smallest in D, the slowest; a verifier should therefore not expect the ratio measured on its development machine to carry to production. For OB3 and DCC-style credentials the mean-median gap and $\sigma$ are large (OB3 in Environment~A: mean 2.067\,ms, $\sigma$ 1.688\,ms), indicating that complex context processing has a long tail.

This result shows that the observation in Section~\ref{sec:breakdown}---that for the simple credential canonicalization is the largest stage but of the same order as the signature---does not generalize to production schemas. This is the most consequential practical finding of our measurements: without any adversarial input, canonicalization of production education schemas is the dominant cost of the pipeline. Canonicalizing one such credential costs six to eleven times what the end-to-end verification step of Table~\ref{tab:e2e} costs, depending on the environment, a step-against-pipeline comparison stated as such above. In domains where W3C VCDM adoption is advancing with education credentials (OB3/DCC), the mitigations that apply are pre-compiled contexts and caching of expansion results. A bound on blank node count, which is the form such a guard usually takes, would not address it: as shown above, context expansion rather than blank node labeling is the dominant term for these schemas. SD-JWT VC and mdoc require no canonicalization, so schema complexity has limited impact on their verification cost.

\section{Discussion}\label{sec:disc}

\subsection{Discussion on Signature Verification Performance}\label{sec:disc-perf}

Across the four environments, mdoc and SD-JWT VC differ in their end-to-end totals by 12\%, 28\%, 19\% and 10\%---the smallest at the edge of our indistinguishability band---while W3C VCDM takes 1.73 to 2.04$\times$ mdoc and 1.57 to 1.79$\times$ SD-JWT VC in every one of them. The asymmetry between those two statements organizes this discussion: one difference is robust to a fivefold change in machine speed and the other is not. The structural reason SD-JWT VC achieves the fastest signing step is that the adoption of JWT Compact Serialization eliminates the need for preprocessing (canonicalization) of the signing target data. The results---signing at p50: 0.038\,ms and verification at p50: 0.117\,ms---reflect the low computational cost of EdDSA (Ed25519) and the preprocessing-free structure. In the Ed25519-unified benchmark (Table~\ref{tab:t13}), SD-JWT VC also signed fastest (p50: 0.040\,ms), confirming a structural advantage in the bare signing step that is independent of algorithm (verification under a unified algorithm was equivalent to mdoc; at issuance with 20 Disclosures, mdoc is cheaper, Table~\ref{tab:e2e}). As the reference measurements in Table~\ref{tab:micro} show, however, routing through the full JWT pipeline of the \texttt{jose} library (\texttt{SignJWT} / \texttt{jwtVerify}, including claims validation) yields signing at 0.101\,ms and verification at 0.174\,ms (p50) in Environment~A---2.6$\times$ and 1.5$\times$ the raw cryptographic operations, and 1.8--2.6$\times$ and 1.2--1.5$\times$ across the four environments. Whether the structural advantage translates into production performance thus depends on the library stack in use.

From the W3C VCDM signing breakdown (Table~\ref{tab:breakdown}), for a simple credential in Environment~A the two RDFC-1.0 canonicalizations that the cryptosuite performs account for 61\% of signing and the Ed25519 step for 27\% (both on a p50 basis); those shares are 63\% and 29\% in B, 57\% and 41\% in C, and 75\% and 21\% in D, so canonicalization is the largest stage in all four while its share is not transferable. That is, as long as the credential is simple (a single blank node), canonicalization overhead remains comparable to the signature operation. This observation does not generalize to complex credentials: in the evaluation with production schemas (Section~\ref{sec:complex}, Table~\ref{tab:t14}), canonicalizing an Open Badges v3.0 credential took p50: 1.603\,ms, about 34$\times$ the simple case and about eight times the whole end-to-end verification of a simple credential. SHA-256 hashing (p50: 0.002\,ms) is not a bottleneck.

Overall, W3C VCDM verification (p50: 0.209\,ms) is 1.8$\times$ that of SD-JWT VC (0.117\,ms) in Environment~A, and 1.40 to 1.90$\times$ across the four. Ratios of this kind are environment-dependent in magnitude: the signing ratio of W3C VCDM to SD-JWT VC is 3.6 in Environment~A, 2.4 in C and 4.4 in D, so Environment~A sits between the extremes rather than at one. One ratio is stable to within 5\% in all four, and it is the one internal to the cryptosuite---two canonicalizations against one, 1.90 to 1.99 with Environment~A at 1.90---which is what we would expect of a quantity that the specification fixes and the machine does not. The source of the gap is therefore the canonicalization the specification requires, not the particular library: every RDFC-1.0 implementation measured for this work, across three libraries and three runtimes, places the canonicalizing configuration last (Section~\ref{sec:lang}), and replacing RDFC-1.0 with JCS in the same Data Integrity proof brings W3C VCDM verification to within 5\% of SD-JWT VC in three environments and within 12\% in the fourth. The configuration with canonicalization removed verifies at 0.119\,ms, but that number is a lower bound on the remaining pipeline rather than evidence about the library: Table~\ref{tab:prim} shows the Ed25519 primitive alone costs 0.116\,ms, so what that row mostly measures is the signature operation both formats share. However, attribute scaling (Table~\ref{tab:t11}: 6.7$\times$ the JCS canonicalization of the same credential at 500 attributes, the comparison between like steps) and the complex credential evaluation (Table~\ref{tab:t14}: about 34$\times$ for the OB3 schema) show that performance degradation with credential size and structure is a format-specific concern.

mdoc requires CBOR encoding/decoding, COSE\_Sign1 structure construction, and per-data-element SHA-256 digest computation. Signing costs p50: 0.112\,ms and verification 0.131\,ms. Once CBOR decoding and per-element digest matching are counted, mdoc is therefore not the fastest verifier at this level: in Environment~A, SD-JWT VC verifies in 0.117\,ms, 12\% less. Having the same measurement in four environments shows that this particular pair has no environment-independent ordering at all. The mdoc-to-SD-JWT VC verification ratio is 1.12 in Environment~A, 1.05 in B, 0.95 in C and 1.80 in D: mdoc is slower in A, indistinguishable in B and C, and much slower in D, which is the machine where Node.js verifies P-256 more slowly than Ed25519 while the other three do the reverse, and where the P-256 signing penalty is largest (the Node.js P-256/Ed25519 signing ratio is 1.86 in D against 1.22 in A). mdoc is the only one of the three formats on P-256 by default, so that one primitive inversion accounts for the whole effect---and unifying the algorithm removes it, bringing the ratio to 0.82--0.98 in all four (Section~\ref{sec:unified-micro}). The gap between these two formats moves with the machine in both direction and magnitude; what does not move is their joint distance from W3C VCDM. Under the algorithm-unified end-to-end comparison (Table~\ref{tab:unified-e2e}), which holds the measured scope fixed, mdoc verification is 0.140\,ms against SD-JWT VC's 0.142\,ms---the gap closes entirely, where in the native configuration mdoc was the slower of the two. What moves between the two configurations is therefore the signature primitive; the structural part of mdoc verification stays where it is. The cross-language results make the same point from the other side: in Go's standard library Ed25519 verification is faster than ECDSA P-256 in all four environments (0.065 against 0.079\,ms in Environment~A), which reverses the ordering again (Section~\ref{sec:lang}). For signing, ECDSA P-256 is the slower primitive in Node.js in every environment (0.046 against 0.038\,ms in Environment~A), and mdoc signing is 2.9$\times$ SD-JWT VC's in Environment~A once MSO construction is included.

Note that in this paper's performance comparison, Ed25519 (EdDSA) was used for SD-JWT VC and W3C VCDM, and ECDSA P-256 was used for mdoc, so the measured results include the effect of signature algorithm differences. The Ed25519-unified experiment (Section~\ref{sec:unified-micro}, Table~\ref{tab:t13}), which also unifies the attribute count, shows that SD-JWT VC signs fastest under the same algorithm, that SD-JWT VC and mdoc verify at comparable speed, and that W3C VCDM remains 1.5$\times$ slower---a residual gap attributable to the verification pipeline (canonicalization/preprocessing) rather than to the algorithm.

What these differences are worth in a deployment depends entirely on what else the verifier does. The spread between the fastest and slowest format is 0.27\,ms per exchange in Environment~A, and 0.14 to 0.71\,ms across the four environments. A verifier that resolves an issuer key over the network, fetches a status list and checks holder binding will spend one to three orders of magnitude more than that on those steps alone, and none of them is measured here (Section~\ref{sec:bench-speed}). The format choice is therefore decisive for throughput in exactly one situation: a verifier that has already eliminated the network from its hot path---cached or statically embedded keys, cached status, no holder-binding round trip---and is CPU-bound on verification. At 0.265 against 0.534\,ms in Environment~A, a single core sustains about 3,800 exchanges per second with mdoc and about 1,900 with W3C VCDM; in Environment~C the same two figures are about 6,700 and 3,500, and in D about 1,300 and 780. The absolute throughput therefore varies by a factor of five across the machines while the factor between the two formats stays near two, so the choice doubles or halves the hardware whatever the machine---which is the form in which this result is usable. Where the verifier is not in that situation---and a proximity presentation over NFC or BLE, whose transport alone takes tens to hundreds of milliseconds, is firmly not---the differences reported here are real but immaterial beside the rest of the exchange.

\subsection{Discussion on Implementation Complexity}

The SD-JWT VC verification reference implementation completes in approximately 10 lines, with only four branch points (token shape, alg allowlist, missing iss, and missing vct). The small code volume is favorable for review, auditing, and security assessment, and the single external dependency (jose) also facilitates supply chain risk management.

W3C VCDM, with 35 lines of code, 4 asynchronous steps, and a cyclomatic complexity of 9, is the most complex of the three wrappers. In particular, resolving the external contexts referenced from @context embeds a network dependency in the verification process. In the default configuration this resolution is an HTTP fetch, but in practice implementations commonly avoid network I/O through caching or statically embedded contexts (our measurements use a static loader), so an external fetch does not necessarily occur on every verification. That choice is also the one the JSON-LD Working Group's Best Practices draft \cite{ref33} asks for: it lists context spoofing and processing delay among the risks of resolving contexts over the network, and recommends that implementations provision vetted local representations of the contexts they expect, resolve only those, and raise an error otherwise. The configuration we measured is therefore the recommended one rather than a favorable choice of ours, though it remains a configuration the deployer has to make. Where no cache is available, or when a cached context expires, network latency, DNS failures, and outages of the context server become direct availability risks, and this dependency itself remains a format-specific property. Its two direct dependencies, \texttt{jsonld} and \texttt{node:crypto}, are the smaller part of the picture: \texttt{jsonld} pulls in \texttt{rdf-canonize} and the context packages transitively, and it is that transitive closure, not the direct count, that bounds the supply chain surface.

mdoc sits in between at 25 lines and 2 asynchronous steps, but its cyclomatic complexity of 8 is close to W3C VCDM's and above SD-JWT VC's 5, because the per-element digest loop contributes branches that the line count does not reveal. The processing separates cleanly into CBOR decoding and COSE\_Sign1 verification, it makes no network calls in any configuration, and its one third-party dependency (\texttt{cbor-x}) carries no network-related risk.

From a developer experience perspective, SD-JWT VC has low learning cost as signing and verification complete with a single node:crypto Ed25519 sign/verify API call. W3C VCDM requires understanding three layers: JSON-LD processor, RDF canonicalization, and cryptographic libraries, presenting a high barrier to entry. mdoc requires understanding CBOR/COSE, but the processing flow is clear in two stages, with intermediate learning cost.

\subsection{Trade-off Structure of Format Characteristics}

The comparison of the three formats reveals the following trade-off structure. SD-JWT VC offers the fastest bare signing step, the simplest implementation, and a mature library ecosystem, but provides no semantic interoperability. VCDM 2.0 + Data Integrity provides semantic interoperability through its JSON-LD representation and RDF dataset canonicalization, but in the configuration of Table~\ref{tab:micro} it is at a disadvantage in both performance and implementation complexity (with canonicalization removed its figures approach SD-JWT VC's, but that configuration is reported only as a lower bound, because the signature primitive alone accounts for most of it---Section~\ref{sec:lang}). Attribute-count scaling (Table~\ref{tab:t11}) confirms that the RDFC-1.0 canonicalization cost of W3C VCDM grows faster as the number of attributes increases, with the per-interval exponent rising monotonically in all four environments (0.49 to 0.96 in Environment~A, and from 0.38--0.52 to 0.83--0.96 across the four), so its use with large credentials requires care. mdoc, the ISO/IEC 18013-5 format for mobile identity documents, combines efficient CBOR binary encoding with a robust security model, but its scope as a general-purpose VC format is limited.

The cross-language comparison (Figures~\ref{fig:f1}--\ref{fig:f3}) shows that the trade-off structure---W3C VCDM being slowest under this paper's implementation conditions---holds in all twelve combinations of three runtimes and four environments. The relative verification ordering of SD-JWT VC and mdoc, by contrast, depends on the cryptographic primitives of the runtime \emph{and} on the machine jointly: mdoc (ECDSA P-256) verifies faster in Python in Environments~A, B and C but not in D, and SD-JWT VC (Ed25519) verifies faster in Go in all four (Section~\ref{sec:lang}). For selective disclosure (Table~\ref{tab:t12}, Fig.~\ref{fig:f5}), W3C VCDM, its JCS variant and mdoc grow with the number of disclosed attributes while SD-JWT VC depends least on the disclosure count; in all cases the cost is on the order of microseconds and is not dominant. The serialization results (Table~\ref{tab:t10}) show that encoding/decoding differences are negligible compared with cryptographic processing, so the serialization format is not a primary decision factor in format selection. As the complex-credential evaluation shows (Table~\ref{tab:t14}), however, schema complexity---context size and blank node count---directly translates into verification cost for W3C VCDM. Finally, the EUDIW is required, through the implementing acts adopted under eIDAS 2.0 \cite{ref8}, to support both mdoc and SD-JWT VC, and the quantitative data in this paper can serve as a reference for implementers evaluating the performance implications of these two mandated formats.

\subsection{What the Measurements Support}

Table~\ref{tab:t15} states, for each situation a deployment is likely to be in, which format costs less and---more importantly---the condition under which that follows from the data reported here. Two situations are deliberately absent because this paper has no evidence on them: proximity presentation over NFC or BLE, where the transport takes tens to hundreds of milliseconds against a format spread of 0.27\,ms and we measured no transport, and attack surface and operational security, which the companion paper treats and which the wrapper metrics of Table~\ref{tab:t8} do not address.

\begin{table*}[!t]
\centering
\caption{What the measurements in this paper support. Each row states the condition under which it holds. Two rows are not performance conclusions and say so: they record requirements that make a format choice before performance enters, and are kept because they are what makes the measured cost unavoidable.}\label{tab:t15}
\footnotesize\setlength{\tabcolsep}{4pt}
\begin{tabular}{@{}>{\raggedright\arraybackslash}p{0.24\textwidth}@{\hspace{5pt}}>{\raggedright\arraybackslash}p{0.12\textwidth}@{\hspace{5pt}}>{\raggedright\arraybackslash}p{0.56\textwidth}@{}}
\toprule
\textbf{Situation} & \textbf{Lower cost} & \textbf{Condition under which this follows from our data}\tabularnewline
\midrule
CPU-bound verification at high throughput & mdoc, then SD-JWT VC & Holds only once key resolution, status checking and holder binding are off the hot path; otherwise those steps dominate (Section~\ref{sec:disc-perf}). Ordering is stable across all four environments and both algorithm choices (Table~\ref{tab:unified-e2e}), though the margin between mdoc and SD-JWT VC is not (Section~\ref{sec:e2e}).\tabularnewline
Selective disclosure with the issuer signature preserved & SD-JWT VC, mdoc & On the Data Integrity side this requires \texttt{ecdsa-sd-2023}, which costs 3.4\,ms to issue and 1.5--2.4\,ms to derive and verify in Environment~A (Table~\ref{tab:sd}), roughly ten times the re-signing path that does not preserve it.\tabularnewline
Education / learner credentials carried as Open Badges v3.0 & W3C VCDM (OB3 profile) & Not a performance conclusion: the 1EdTech profile is defined in JSON-LD, so the format follows from the ecosystem. Our contribution here is the cost to budget for---canonicalization at 34$\times$ a simple credential (Table~\ref{tab:t14})---and the mitigation that applies, which is context pre-compilation rather than a blank node limit (Section~\ref{sec:complex}).\tabularnewline
Semantic interoperability across vocabularies & W3C VCDM & Not a performance conclusion; stated for completeness because it is the requirement that makes the canonicalization cost unavoidable.\tabularnewline
\bottomrule
\end{tabular}
\end{table*}

As a note on the education domain: 1EdTech Open Badges v3.0 \cite{ref34} and the Comprehensive Learner Record (CLR) \cite{ref35} are defined as VCDM-conformant JSON-LD credentials, and interoperability efforts such as the DCC \cite{ref36} and the JFF Plugfest \cite{ref31}---the interoperability events hosted by Jobs for the Future, described in Section~\ref{sec:bench-canon}---adopt the same profile. W3C VCDM is therefore the de facto standard for education credentials; however, as shown in Section~\ref{sec:complex}, the canonicalization cost of the OB3 schema reaches about 34$\times$ that of a simple credential, and Section~\ref{sec:complex} attributes that cost to context expansion rather than to blank node count. Verifiers performing bulk verification should therefore reach for statically embedded contexts and caching of expansion results. Carrying the same claims in an enveloping-proof (JOSE/COSE) profile would avoid canonicalization altogether, but we did not measure that configuration and offer it as a direction rather than as a recommendation this paper supports.

\section{Limitations and Future Work}\label{sec:limits}

This study has several limitations. First, Environment~A has SMT disabled and unused cores taken offline, but it is a cloud VM rather than a physically dedicated host, so hypervisor-level effects cannot be fully excluded; core pinning, which would be the usual remedy, made the run-to-run spread worse there rather than better (Section~\ref{sec:env}). Environments~B and D are bare metal---D with core pinning, B with the clock held at its base frequency---and they reproduce the ordering, which is what the four-environment design buys here. All four share the same Node.js build, Go toolchain and library versions, so a difference between any two of them is attributable to hardware; the one version difference left is Python, 3.12.3 in A against 3.12.15 in the other three. Environments~A, B and C do not pin the measurement process to a core, for the different reasons given in Section~\ref{sec:env}. Our warmup is fixed at 50 iterations and we do not test whether the virtual machine has reached a steady state, which for JIT-compiled virtual machines is known not to be guaranteed \cite{ref37}; the run-to-run spreads we report bound the consequence without excluding it. We mitigate this by reporting medians of statistics across five independent runs, reporting outlier counts, and interpreting results primarily through p50. JIT optimization and garbage collection effects remain in all four environments, so replication with an implementation in a systems language such as C or Rust is still desirable.

Second, although the cross-language comparison (Figs.~\ref{fig:f1}--\ref{fig:f3}) was performed on the same machine within each environment, each language's runtime characteristics (Go: native binary, Python: CPython interpreter, Node.js: V8 JIT) and cryptographic library implementations (OpenSSL-backed vs. Go standard library) differ, so absolute cross-language differences must be interpreted with care. The comparison targets the relative ordering among formats and the trends.

Third, the complex-credential evaluation (Section~\ref{sec:complex}) is limited to two Open Badges v3.0-family schemas and synthetic credentials; extending it to other production schemas---multi-namespace mDL mdocs, EUDIW (eIDAS 2.0) PID/EAA schemas, and so on---remains future work.

This paper additionally evaluated attribute-count scaling (Section~\ref{sec:scaling}), selective disclosure (Section~\ref{sec:seldisc}), the Ed25519-unified benchmark (Section~\ref{sec:unified-micro}), and canonicalization of complex credentials (Section~\ref{sec:complex}). Future work includes measurement in browser/WebCrypto environments, memory-usage measurement, and extension to additional languages and runtimes such as Rust and Java.

\section{Threats to Validity}\label{sec:ttv}

We discuss threats to validity that may affect the generalizability of our results.

\textit{Internal validity.} Benchmarks were measured with \texttt{process.hrtime.bigint()} on the Node.js V8 runtime, so JIT optimization stages, GC pauses, and event-loop interference may affect the results. After 50 warmup iterations we measure N=2,000 iterations per run and report, for each statistic, its median across five independent runs; outliers (Tukey 1.5$\times$IQR) are reported without removal, and p50---robust to outliers---is used as the representative statistic. GC effects nonetheless cannot be entirely eliminated. Measurements were performed on hosts reserved for measurement, with SMT disabled where the CPU has it, and with the process pinned to a single core in Environment~D. Environment~A is a cloud VM rather than a physically dedicated machine, so hypervisor-level effects cannot be fully excluded; the two bare-metal environments reproduce the ordering and the structure of the breakdown (Section~\ref{sec:e2e}). Environments~A, B and C do not pin the process---on A because pinning increased the spread, on B because the clock is instead held at its base frequency by fixing the governor and disabling turbo, and on C because macOS provides no equivalent---so only Environment~D is pinned, and we report the isolation of each environment as it was rather than as uniform.

\textit{Dependence on library versions.} The absolute values reported here are tied to the library versions in Table~\ref{tab:t5}, and that dependence is not always small. A separate measurement on the Environment~C machine, under this paper's own protocol (N=2,000 $\times$ 5 processes), found that Python Ed25519 signing is 3.28$\times$ slower and verification 2.42--2.45$\times$ slower under \texttt{cryptography} 49.0.0 than under 50.0.2, with the Python interpreter version (3.12 against 3.14) making a 0--1.1\% difference; ECDSA P-256 moved by 1.29$\times$ for signing and 1.12$\times$ for verification between the same two releases. The two releases differ in major version and in the OpenSSL they bundle (4.0.1 against 4.0.3), so the change is not a patch-level one on the Python package, whatever it is on the library beneath. On that machine the effect is large enough to reorder a figure: under 49.0.0 the fastest Python signer is mdoc rather than SD-JWT VC. The same comparison on the Environment~D machine showed no difference at all between the two releases---ratios of 0.98 to 1.00 for every primitive, in two runs and under both interpreters---so the effect is one we observe on the arm64 machine and do not observe on that older x86\_64 host; whether newer x86\_64 hardware behaves like either is not something these measurements settle. We measured with \texttt{cryptography} 50.0.2 in all four environments and record the version with every result file, but readers comparing against their own numbers should check the version before attributing a difference to anything else. This is also why the cryptographic primitive baseline of Table~\ref{tab:prim} is reported: it localizes such a change to the primitive instead of leaving it to contaminate a format comparison.

\textit{Construct validity.} The benchmarks use Ed25519 for SD-JWT VC and W3C VCDM and ECDSA P-256 for mdoc, so differences in the verification pipeline (canonicalization and preprocessing) are confounded with differences in signature algorithm. This reflects the representative algorithm choice of each ecosystem, but a strict evaluation of format-specific differences requires a comparison in which all formats use the same algorithm. The Ed25519-unified benchmark (Table~\ref{tab:t13}) partially isolates this confound and confirms that SD-JWT VC has the fastest bare signing step independently of the algorithm. Conversely, in the equal-scope comparison of Table~\ref{tab:unified-e2e}, mdoc verification in Environment~A is 0.140\,ms under Ed25519 against 0.127\,ms under P-256, so the substitution costs mdoc rather than helping it---consistent with the primitive baseline, where Node.js verifies P-256 faster than Ed25519 there. In Environment~D, where that primitive ordering is reversed, so is the effect of the substitution, which is the clearest evidence we have that this particular comparison is about primitives rather than formats. This is consistent with the cross-language comparison in Section~\ref{sec:lang}. The serialization results (Table~\ref{tab:t10}) and attribute-count scaling (Table~\ref{tab:t11}) also contribute to isolating format-specific overhead.

\textit{Statistical conclusion validity.} The per-iteration $\sigma$ and percentiles in the tables describe the distribution within a single run; iterations are autocorrelated through JIT compilation and garbage collection and are not independent samples, so those columns cannot support a claim that two conditions differ \cite{ref38,ref39}. Claims of difference in this paper rest on the five independent per-run p50 values, whose spread the end-to-end table reports, and on the 10\% threshold defined in Section~\ref{sec:env}. Five runs is few: it is enough to show that the orderings we report are stable and not enough to put a confidence interval on a 12\% gap, and we have not claimed one. Having four environments does not substitute for more runs within one---they are not independent replicates of the same quantity, since each measures a different machine---but it does bound what a single machine could have told us: a difference that survives a fivefold change in absolute speed, as the W3C VCDM gap does, is not an artifact of one host, whereas a difference that changes sign across the four, as the mdoc--SD-JWT VC gap at the verification step does, could not have been identified as fragile from one environment at all.

\textit{External validity.} The credentials used in the main signing/verification benchmarks (Tables~\ref{tab:e2e} and~\ref{tab:micro}) have a simple structure composed of a few fields such as name, date of birth, and address. Attribute-count scaling from 5 to 500 attributes is evaluated in Section~\ref{sec:scaling} (Table~\ref{tab:t11}), and canonicalization of complex, nested schemas containing blank nodes is evaluated in Section~\ref{sec:complex} (Table~\ref{tab:t14}) with Open Badges v3.0, DCC-style, and synthetic credentials; the complex schemas nevertheless center on the education-domain OB3 profile, and generalization to production schemas in other domains requires further experiments. The simple credential also embeds its \texttt{@context} rather than referencing the W3C context document (Section~\ref{sec:bench-speed}), so its canonicalization figures are a lower bound for a conforming VCDM 2.0 credential, which carries \texttt{https://www.w3.org/ns/credentials/v2} and expands it. The measurements were performed with Node.js, Go and Python in the four environments of Table~\ref{tab:t4}---three x86\_64 machines spanning a cloud VM, a low-power bare-metal host and a server of around 2010 without AVX, plus one arm64 desktop machine---and the cross-language results are limited to relative ordering and trends. What those four support, and what they do not, divides cleanly. The ordering between formats and the structure of the breakdown reproduce in all four. The absolute values do not: relative to Environment~A, the whole kit runs 0.43--0.80$\times$ on the arm64 machine (median 0.63 over the 118 conditions above 0.01\,ms; below that the rounding of the reported p50 dominates the ratio) and 1.81--3.95$\times$ on the old x86\_64 host (median 2.54). Neither factor is uniform across kinds of work: on the arm64 machine rows containing canonicalization are 1.3--2.2$\times$ faster while rows dominated by a cryptographic primitive are only 1.2--1.6$\times$ faster, and on the old x86\_64 host it is the P-256 rows that reach the top of the range (2.5--3.9$\times$, against 1.9--3.7$\times$ for the rows containing canonicalization). A single scaling constant therefore does not carry results between these machines, which is why we quote none. Because the factor is not uniform, ratios between formats move in both directions around Environment~A, and we report every ratio with its environment attached. Four machines is still four: browser and WebCrypto environments, mobile hardware, and other architectures may behave differently, and nothing here establishes a trend in hardware generation from a sample of this size.

\textit{Validity of the complexity analysis.} Lines of code depend heavily on implementation style, the presence of error handling, and the degree of library use. Moreover, as stated in Section~\ref{sec:bench-speed}, the functional scope of the reference implementations is not identical across formats: the SD-JWT VC reference implementation does not include Disclosure matching, and the W3C VCDM reference implementation canonicalizes the document only, without the proof options (Listing 14). The complexity results are therefore a comparison within the reference implementations and metrics of this paper and do not assert an ordering of the inherent complexity of the formats. For performance, Table~\ref{tab:e2e} reports only the comparison with the functional scope equalized, and the step-level numbers are positioned as a decomposition of that total. Cyclomatic complexity is computed over the entire deserialization function, so the value varies with function decomposition granularity. The number of external dependencies is a proxy for supply chain risk but does not account for library maturity, CVE history, transitive dependencies, or maintenance status.

\section{Conclusion}

This paper conducted a reproducible comparative evaluation of SD-JWT VC, W3C VCDM, and mdoc along two axes---signature verification performance and implementation complexity---and provided quantitative data that implementers and standardization stakeholders can consult when selecting a format.

For performance, measurement in four execution environments (medians across N=2,000 $\times$ 5 runs, every benchmark in every environment) covered all three formats over an identical end-to-end scope: issue a credential, present it with 5 of 20 attributes disclosed, and verify the disclosed claims. The end-to-end totals in Environment~A are 0.265\,ms for mdoc, 0.298\,ms for SD-JWT VC and 0.534\,ms for W3C VCDM. The four environments run the same Node.js build, Go toolchain and library versions, span about a factor of five in absolute speed, and separate the robust result from the fragile one. W3C VCDM takes 1.73--2.04$\times$ mdoc and 1.57--1.79$\times$ SD-JWT VC in all four, a margin at least five times the 10\% indistinguishability band, and the ordering also survives repeating the scenario with Ed25519 for all three formats and again with P-256 for all three---so it is a property of the pipelines rather than of the machine or of the algorithm pairing. The difference between mdoc and SD-JWT VC is not of that kind: it is 12\%, 28\%, 19\% and 10\% in the four environments, the smallest at the edge of our indistinguishability band, and at the verification step the order flips in Environment~D, where SD-JWT VC verifies faster than mdoc (0.315 against 0.377\,ms, Table~\ref{tab:e2e}); mdoc is the only format on P-256 by default, and which of P-256 and Ed25519 verifies faster depends on the runtime's cryptographic implementation and on the machine (Table~\ref{tab:prim}). The W3C VCDM gap is produced by RDFC-1.0 canonicalization, which \texttt{eddsa-rdfc-2022} performs twice per operation---over the document and over the proof options---and which therefore runs six times in one end-to-end exchange---twice each in issuance, presentation and verification. That attribution rests first on replacing RDFC-1.0 with JCS, which brings W3C VCDM verification to within 5\% of SD-JWT VC in three environments and within 12\% in the fourth, and on the two-canonicalizations-to-one ratio coming out at 1.90--1.99 in all four environments; three RDFC-1.0 implementations in three language runtimes also place W3C VCDM last in all twelve runtime-environment combinations, which is the weakest of the three because those implementations share a lineage. It does not rest on the canonicalization-free configuration, which we report only as a lower bound because the signature primitive alone accounts for most of it. One qualification applies to the comparison itself: \texttt{eddsa-rdfc-2022} has no derivation step, so its presentation re-signs the disclosed subset and the verifier checks the holder's signature rather than the issuer's. Measuring \texttt{ecdsa-sd-2023}, which does preserve the issuer signature, puts issuance at 3.4\,ms in Environment~A and derivation and verification at 1.5--2.4\,ms---issuance alone is 15 to 21 times the \texttt{eddsa-rdfc-2022} issuance it replaces, depending on the environment---so the end-to-end ratio is a lower bound on the cost of W3C VCDM at equal guarantee. With production education schemas, canonicalization of an Open Badges v3.0 credential costs 1.603\,ms in Environment~A, 31--49$\times$ a simple credential across the four, and blank node labeling is not the dominant term in any of them---the cost tracks context expansion, though our inputs do not separate the two.

In format selection, the requirements of the target use case---interoperability, performance, and regulatory compliance---should be evaluated comprehensively, and the trade-offs should be accepted explicitly. The security characteristics of the verification pipelines that this paper does not address---canonicalization-induced DoS, context injection, SSRF, and so on---are analyzed in a companion paper by the authors \cite{ref40} that uses the same measurement infrastructure. Our evaluation framework and benchmark tool are published as open source \cite{ref28}, and we hope they support replication under different environments and conditions.

\section*{Data Availability}

The raw result files behind every number in this paper are deposited at \url{https://doi.org/10.5281/zenodo.23933321} under CC BY 4.0. The deposit holds eight runs---the four environments of Table~\ref{tab:t4}, each measured once with the performance suites and once with the attack-surface suites of the companion paper \cite{ref40}. Each run carries its own \texttt{environment.txt} recording the host, the isolation settings and the kit commit, a \texttt{summary.json} giving every condition's runs, n, mean, standard deviation, 95\% confidence interval and percentiles, and the five independent per-run outputs those statistics are computed from. The accompanying README maps runs to environments and gives the \texttt{jq} commands that read the figures reported here back out of the archive. The measurement kit that produced the files is separate and is cited as \cite{ref28}.

\section*{Appendix A: Implementation Code}

This appendix presents excerpts from the measurement implementations. Listings 7, 9, 11 and 12 are the per-format signing and verification benchmarks of Table~\ref{tab:micro}; Listing 16 is the end-to-end implementation behind Table~\ref{tab:e2e}, which is a different code path and is reproduced here because the headline totals come from it. All are excerpts from the Node.js engine of the measurement kit (vcbench, \texttt{node/bench.mjs}, JavaScript/ESM), executed with Node.js v24.18.0 (Table~\ref{tab:t4}). Two properties of the kit are worth noting: every benchmark checks the result of the operation it times and raises on failure, and the Data Integrity listings canonicalize both the document and the proof options, as the cryptosuite requires. Each benchmark runs through the shared \texttt{bench}/\texttt{benchAsync} helper (pseudocode in Listing 6), which records each iteration individually with \texttt{process.hrtime.bigint()} after warmup. The routines in Listings 8 and 10, and the annotated versions in Appendix A.4, are the TypeScript verification reference implementations analyzed in the complexity study (Section~\ref{sec:bench-cplx}, Table~\ref{tab:t8}).

\subsection*{A.1 SD-JWT VC Signing and Verification}

As documented in Table~\ref{tab:t5}, SD-JWT VC signing and verification are completed with node:crypto sign / verify calls (the full jose JWT pipeline is measured separately as a reference). Listing 7 shows the implementation used for measurement, and Listing 8 the deserialization reference implementation used for the complexity analysis.

\Needspace*{5\baselineskip}
\textbf{Listing 7. SD-JWT VC Benchmark Implementation (vcbench/node/bench.mjs)}\nopagebreak

\begin{lstlisting}
// vcbench/node/bench.mjs --- SD-JWT VC (node:crypto)
async function runSdJwt() {
  const { privateKey, publicKey } =
  crypto.generateKeyPairSync('ed25519')
  const header = b64url(Buffer.from(JSON.stringify(
        { alg: 'EdDSA' })))
  const payload = b64url(Buffer.from(JSON.stringify({
          iss: 'https://issuer.example.com',
          vct: 'identity', sub: 'did:example:holder' })))
  const sigInput = `${header}.${payload}`
  // Sign: Ed25519 over the JWS Signing Input + token assembly
  bench('sdjwt/stdcrypto/sign', N, () => {
      const s = crypto.sign(null, Buffer.from(sigInput), privateKey)
      void `${sigInput}.${b64url(s)}`
  })
  // Verify: token split + Ed25519 verification
  const token = `${sigInput}.` +
  b64url(crypto.sign(null, Buffer.from(sigInput), privateKey))
  bench('sdjwt/stdcrypto/verify', N, () => {
      const p = token.split('.')
      crypto.verify(null, Buffer.from(`${p[0]}.${p[1]}`),
        publicKey, Buffer.from(p[2], 'base64url'))
  })
}
\end{lstlisting}

\Needspace*{5\baselineskip}
\textbf{Listing 8. SD-JWT VC Verification Reference Implementation (as analyzed in the complexity study)}\nopagebreak

\begin{lstlisting}
async function deserializeSdJwt(
  token: string, publicKey: CryptoKey
) {
  const parts = token.split('.')
  if (parts.length !== 3) throw new Error('Invalid JWT')
  // alg allowlist verification
  const header = JSON.parse(atob(parts[0]))
  if (!['ES256','ES384','EdDSA','RS256'].includes(header.alg))
  throw new Error('Unsupported algorithm')
  // Signature verification + claim retrieval (1 API call)
  const { payload } = await jwtVerify(token, publicKey)
  // Required claims check
  if (!payload.iss) throw new Error('Missing issuer')
  if (!payload.vct) throw new Error('Missing vct')
  return payload
}
\end{lstlisting}

\subsection*{A.2 W3C VCDM Signing and Verification}

The signing process for W3C VCDM consists of three stages: RDFC-1.0 canonicalization, SHA-256 hashing, and Ed25519 signing. Listing 9 shows the implementation used for measurement: the jsonld library performs canonicalization, and node:crypto performs hashing and signing. The breakdown in Table~\ref{tab:breakdown} was obtained by measuring each stage individually (canonicalization only, hashing only, signing only) with the same bench helper.

\Needspace*{5\baselineskip}
\textbf{Listing 9. W3C VCDM Benchmark Implementation (vcbench/node/bench.mjs)}\nopagebreak

\begin{lstlisting}
// W3C VCDM 2.0 + Data Integrity (eddsa-rdfc-2022).
// The cryptosuite canonicalizes the document AND the
// proof options and signs the 64-byte concatenation
// of their digests; measuring one canonicalization
// understates the suite.
const RDFC_OPTS = { algorithm: 'URDNA2015',
  format: 'application/n-quads', safe: false }
const sha256b = (b) =>
  crypto.createHash('sha256').update(b).digest()

const proofOptions = diProofOptions('eddsa-rdfc-2022')
async function diHashData(jsonld, doc, proofOptions) {
  const [proofNQ, docNQ] = await Promise.all([
    jsonld.normalize(proofOptions, RDFC_OPTS),
    jsonld.normalize(doc, RDFC_OPTS),
  ])
  return Buffer.concat([sha256b(Buffer.from(proofNQ)),
                        sha256b(Buffer.from(docNQ))])
}
const hashData = () => diHashData(jsonld, VC_DOC, proofOptions)

await benchAsync('jsonld/jsonld-lib/sign', N, async () => {
  crypto.sign(null, await hashData(), privateKey)
})
const sig0 = crypto.sign(null, await hashData(), privateKey)
await benchAsync('jsonld/jsonld-lib/verify', N, async () => {
  if (!crypto.verify(null, await hashData(), publicKey, sig0))
    throw new Error('verify failed')
})
// the two canonicalizations, reported separately
await benchAsync('jsonld/jsonld-lib/normalize-only', N,
  async () => { await jsonld.normalize(VC_DOC, RDFC_OPTS) })
await benchAsync('jsonld/jsonld-lib/normalize-both', N,
  async () => {
    await jsonld.normalize(proofOptions, RDFC_OPTS)
    await jsonld.normalize(VC_DOC, RDFC_OPTS)
  })
\end{lstlisting}

\Needspace*{5\baselineskip}
\textbf{Listing 10. W3C VCDM Verification Reference Implementation (as analyzed in the complexity study)}\nopagebreak

\begin{lstlisting}
async function deserializeJsonLdVc(
  document: Record<string, unknown>,
  signature: Uint8Array, publicKey: Uint8Array
) {
  const loader = makeStaticContextLoader()
  // 1. @context verification
  const ctx = document['@context'] as string[]
  if (!ctx.includes(VC_CONTEXT_URL))
  throw new Error('Missing VC context')
  // 2. Remove proof field
  const { proof, ...docWithoutProof } = document
  // 3. JSON-LD expansion (network retrieval occurs)
  await jsonld.expand(docWithoutProof,
    { documentLoader: loader, safe: false })
  // 4. RDFC-1.0 RDF canonicalization
  // Blank node identification = graph isomorphism problem
  const normalized = await jsonld.normalize(
    docWithoutProof, {
      algorithm: 'URDNA2015',
      format: 'application/n-quads',
      documentLoader: loader, safe: false
  }) as string
  // 5. SHA-256 hash
  const hash = await sha256(normalized)
  // 6. Ed25519 signature verification
  const valid = await ed25519Verify(
    signature, hash, publicKey)
  if (!valid) throw new Error('Signature failed')
  // 7. VC schema verification
  if (!document['credentialSubject'])
  throw new Error('Missing credentialSubject')
  return document
}
\end{lstlisting}

\subsection*{A.3 mdoc (ISO 18013-5) Signing and Verification}

The mdoc measurement implementation consists of CBOR encoding with cbor-x, MSO construction from per-data-element SHA-256 digests, and ECDSA P-256 signing over the COSE\_Sign1 Sig\_Structure (raw r\textbar\textbar s as used by COSE, dsaEncoding: ieee-p1363). Listing 11 shows the signing side and Listing 12 the verification side of the measurement implementation. The MSO built here omits \texttt{deviceKeyInfo} and the tag-24 wrapping of the CDDL in Listings 4 and 5, which the scenario does not exercise. For the full verification reference implementation including per-element digest matching, see the complexity-analysis code (Listing 15 in Appendix A.4).

\Needspace*{5\baselineskip}
\textbf{Listing 11. mdoc Signing-side Measurement Implementation (vcbench/node/bench.mjs)}\nopagebreak

\begin{lstlisting}
// mdoc signing. IssuerSignedItem carries a random salt
// of at least 16 bytes (ISO/IEC 18013-5 9.1.2.5) so that
// an undisclosed element cannot be recovered from its
// digest; the MSO carries version, digestAlgorithm and
// validityInfo.
const MDOC_NS = 'org.iso.18013.5.1'
const buildIssued = () => {
  const items = MDOC_FIELDS.map(([k, v], i) => ({
    digestID: i, random: crypto.randomBytes(16),
    elementIdentifier: k, elementValue: v,
  }))
  const encoded = items.map((it) => cborEncode(it))
  const digests = Object.fromEntries(
    items.map((it, i) => [it.digestID, sha256b(encoded[i])]))
  const mso = cborEncode({
    version: '1.0', digestAlgorithm: 'SHA-256',
    docType: `${MDOC_NS}.mDL`,
    valueDigests: { [MDOC_NS]: digests },
    validityInfo: { signed: '2024-01-01T00:00:00Z',
      validFrom: '2024-01-01T00:00:00Z',
      validUntil: '2029-01-01T00:00:00Z' },
  })
  const protectedHdr = cborEncode({ 1: -7 })
  const sigStruct = cborEncode(
    ['Signature1', protectedHdr, Buffer.alloc(0), mso])
  const sig = crypto.sign('sha256', sigStruct,
    { key: privateKey, dsaEncoding: 'ieee-p1363' })
  return cborEncode({
    docType: `${MDOC_NS}.mDL`,
    issuerSigned: {
      nameSpaces: { [MDOC_NS]: encoded },
      issuerAuth: [protectedHdr, {}, mso, sig],
    },
  })
}
bench('mdoc/cbor-x/sign', N, () => { buildIssued() })
\end{lstlisting}

\Needspace*{5\baselineskip}
\textbf{Listing 12. mdoc Verification-side Measurement Implementation (vcbench/node/bench.mjs)}\nopagebreak

\begin{lstlisting}
// mdoc verification, ISO/IEC 18013-5 9.3.1: decode,
// verify COSE_Sign1 over the reconstructed Sig_structure,
// then match each element's digest against the MSO BY
// digestID. Matching by position breaks on a disclosed
// subset.
const verifyMdoc = (bytes) => {
  const doc = cborDecode(bytes)
  const [protectedHdr, , mso, sig] = doc.issuerSigned.issuerAuth
  const sigStruct = cborEncode(
    ['Signature1', protectedHdr, Buffer.alloc(0), mso])
  if (!crypto.verify('sha256', sigStruct,
      { key: publicKey, dsaEncoding: 'ieee-p1363' }, sig)) {
    throw new Error('COSE_Sign1 signature verification failed')
  }
  const digests = cborDecode(mso).valueDigests[MDOC_NS]
  for (const enc of doc.issuerSigned.nameSpaces[MDOC_NS]) {
    const it = cborDecode(enc)
    if (!Buffer.from(digests[it.digestID])
         .equals(sha256b(Buffer.from(enc)))) {
      throw new Error('digest mismatch')
    }
  }
  return true
}
const issued = buildIssued()
bench('mdoc/cbor-x/verify', N, () => { verifyMdoc(issued) })
\end{lstlisting}

\subsection*{A.4 Code Analyzed in the Complexity Study (Annotated)}

This section lists the verification reference implementations that the complexity metrics in Section~\ref{sec:cplx} (Table~\ref{tab:t8}) were computed on. The branch, asynchronous-step, and network annotations in the comments correspond to the metrics in Table~\ref{tab:t8}.

\Needspace*{5\baselineskip}
\textbf{Listing 13. SD-JWT VC Verification Reference Implementation --- With Complexity Annotations}\nopagebreak

\begin{lstlisting}
// SD-JWT VC deserialization (~10 LOC)
// Direct deps: jose (1)
// Network calls: 0
async function deserializeSdJwt(
  token: string, publicKey: CryptoKey
) {
  // -\/-\/- Preprocessing: Split JWT compact representation -\/-\/-
  const parts = token.split('.')
  if (parts.length !== 3) // (*) Branch 1
  throw new Error('Invalid JWT format')
  // -\/-\/- Header verification -\/-\/-
  const header = JSON.parse(
    atob(parts[0].replace(/-/g,'+').replace(/_/g,'/')))
  if (!['ES256','ES384','EdDSA','RS256']
    .includes(header.alg)) // (*) Branch 2
  throw new Error('Unsupported algorithm')
  // -\/-\/- Signature verification (only async step) -\/-\/-
  const { payload } =
  await jwtVerify(token, publicKey) // (*) async 1
  // -\/-\/- Required claims check -\/-\/-
  if (!payload.iss) // (*) Branch 3a
  throw new Error('Missing issuer')
  if (!payload.vct) // (*) Branch 3b
  throw new Error('Missing vct claim')
  return payload
}
\end{lstlisting}

\Needspace*{5\baselineskip}
\textbf{Listing 14. W3C VCDM Verification Reference Implementation --- With Complexity Annotations}\nopagebreak

\begin{lstlisting}
// W3C VCDM deserialization (~35 LOC)
// Direct deps: jsonld, node:crypto (2)
// Network calls: 0 as measured (static loader);
// 2 at the library's default loader
async function deserializeJsonLdVc(
  document: Record<string, unknown>,
  signature: Uint8Array,
  publicKey: Uint8Array
) {
  // -\/-\/- Document Loader construction -\/-\/-
  // (*) SSRF RISK: allowlist required in production
  const loader = makeStaticContextLoader()
  // -\/-\/- 1. @context verification -\/-\/-
  const ctx = document['@context'] as string[]
  if (!ctx || // (*) Branch 1
    !ctx.includes(VC_CONTEXT_URL))// (*) Branch 2
  throw new Error('Missing VC context')
  // -\/-\/- 2. Separate proof field -\/-\/-
  const { proof: _proof,
    ...documentWithoutProof } = document
  // -\/-\/- 3. JSON-LD expansion -\/-\/-
  // (*) NETWORK 1: Context URL retrieval
  await jsonld.expand( // (*) async 1
    documentWithoutProof,
    { documentLoader: loader,
      safe: false })
  // -\/-\/- 4. RDFC-1.0 RDF canonicalization -\/-\/-
  // (*) DoS RISK: Blank node identification =
  // graph isomorphism -> worst-case exponential
  // (*) NETWORK 2: Additional context retrieval
  const normalized =
  await jsonld.normalize( // (*) async 2
    documentWithoutProof, {
      algorithm: 'URDNA2015',
      format: 'application/n-quads',
      documentLoader: loader,
      safe: false,
  }) as string
  if (!normalized) // (*) Branch 3
  throw new Error('Empty normalization')
  // -\/-\/- 5. SHA-256 hash -\/-\/-
  const hash =
  await sha256(normalized) // (*) async 3
  // -\/-\/- 6. Ed25519 signature verification -\/-\/-
  const valid = await ed25519Verify(// (*) async 4
    signature, hash, publicKey)
  if (!valid) // (*) Branch 4
  throw new Error('Signature failed')
  // -\/-\/- 7. VC schema verification -\/-\/-
  if (!document.type) // (*) Branch 5
  throw new Error('Missing type')
  if (!document.issuer) // (*) Branch 6
  throw new Error('Missing issuer')
  if (!document.credentialSubject) // (*) Branch 7
  throw new Error('Missing subject')
  if (!document.issuanceDate) // (*) Branch 8
  throw new Error('Missing issuanceDate')
  return document
}
\end{lstlisting}

\Needspace*{5\baselineskip}
\textbf{Listing 15. mdoc Verification Reference Implementation --- With Complexity Annotations}\nopagebreak

\begin{lstlisting}
// mdoc deserialization (~25 LOC)
// Direct deps: cbor-x, node:crypto (2)
// Network calls: 0
export async function verifyMdoc(
  mdocBytes: Uint8Array,
  publicKey: CryptoKey
): Promise<boolean> {
  // -\/-\/- 1. CBOR decode -\/-\/-
  const doc = decode(mdocBytes)
  if (!doc?.issuerSigned) // (*) Branch 1
  throw new Error('Invalid mdoc')
  const { issuerAuth, nameSpaces }
  = doc.issuerSigned
  const [protectedHeader, ,
    msoPayload, signature] = issuerAuth
  // -\/-\/- 2. COSE algorithm verification -\/-\/-
  const alg = decode(protectedHeader).get(1)
  if (alg !== -7) // (*) Branch 2
  throw new Error('Unexpected alg')
  // -\/-\/- 3. COSE_Sign1 signature verification -\/-\/-
  const sigStructure = buildSigStructure(
    protectedHeader, msoPayload)
  const valid =
  await crypto.subtle.verify( // (*) async 1
    { name: 'ECDSA', hash: 'SHA-256' },
    publicKey, signature, sigStructure)
  if (!valid) // (*) Branch 3
  return false
  // -\/-\/- 4. MSO decode + digest verification -\/-\/-
  const mso = decode(msoPayload)
  const storedDigests =
  mso.valueDigests['org.iso.18013.5.1']
  const items =
  nameSpaces['org.iso.18013.5.1']
  for (let i = 0; i < items.length; i++) {
    const item = decode(items[i])
    const computed = new Uint8Array(
      await crypto.subtle.digest( // (*) async 2
        'SHA-256', items[i]))
    const expected = storedDigests[item.digestID]
    if (expected.length !== // (*) Branch 4
      computed.length) return false
    for (let j = 0; j < computed.length; j++)
    if (computed[j] !== // (*) Branch 5
      expected[j]) return false
  }
  return true
}
\end{lstlisting}

\subsection*{A.5 End-to-End Implementation (the basis of Table~\ref{tab:e2e})}

The listings above measure signing and verification of a single credential. The end-to-end totals come from a separate code path, excerpted here. Note the three properties that the per-format listings do not show: the Data Integrity path canonicalizes the proof options as well as the document, the mdoc path decodes the CBOR and matches digests by \texttt{digestID}, and every operation's result is checked.

\Needspace*{5\baselineskip}
\textbf{Listing 16. End-to-End Benchmark, abbreviated (vcbench/node/bench.mjs, \texttt{e2e} suite)}\nopagebreak

\begin{lstlisting}
// Data Integrity: the signing input is
// sha256(proofOptions) || sha256(document)
async function diHashData(jsonld, doc, proofOptions) {
  const [proofNQ, docNQ] = await Promise.all([
    jsonld.normalize(proofOptions, RDFC_OPTS),
    jsonld.normalize(doc, RDFC_OPTS),
  ])
  return Buffer.concat([sha256b(Buffer.from(proofNQ)),
                        sha256b(Buffer.from(docNQ))])
}

// A bundles sign/verify for the algorithm under test;
// fullDoc / subsetDoc are the 20- and 5-attribute documents.
// --- W3C VCDM: present = re-canonicalize + re-sign the
// subset, so the verifier checks the HOLDER's signature
const issue = async () =>
  A.sign(await diHashData(jsonld, fullDoc, proofOptions))
const present = async () =>
  ({ sig: A.sign(
       await diHashData(jsonld, subsetDoc, proofOptions)) })
const verify = async (pres) => {
  if (!A.verify(
        await diHashData(jsonld, subsetDoc, proofOptions),
        pres.sig)) throw new Error('verify failed')
  return true
}

// --- SD-JWT VC: verify = JWS check + Disclosure digests
const verifySdJwt = (vp) => {
  const [token, ...disclosures] = vp.split('~').filter(Boolean)
  const [h, pl, sg] = token.split('.')
  if (!A.verify(Buffer.from(`${h}.${pl}`),
                Buffer.from(sg, 'base64url')))
    throw new Error('bad signature')
  const payload =
    JSON.parse(Buffer.from(pl, 'base64url').toString())
  const set = new Set(payload._sd)
  for (const d of disclosures) {
    if (!set.has(b64url(sha256b(Buffer.from(d)))))
      throw new Error('digest mismatch')
  }
  return true
}

// --- mdoc: verify = CBOR decode + COSE_Sign1 +
// per-element digest match keyed by digestID
const verifyMdoc = (vpBytes) => {
  const doc = cborDecode(vpBytes)
  const [protectedHdr, , mso, sig] = doc.issuerSigned.issuerAuth
  const sigStruct = cborEncode(
    ['Signature1', protectedHdr, Buffer.alloc(0), mso])
  if (!A.verify(sigStruct, sig))
    throw new Error('bad signature')
  const digests = cborDecode(mso).valueDigests[MDOC_NS]
  for (const enc of doc.issuerSigned.nameSpaces[MDOC_NS]) {
    const it = cborDecode(enc)
    if (!Buffer.from(digests[it.digestID])
         .equals(sha256b(Buffer.from(enc))))
      throw new Error('digest mismatch')
  }
  return true
}

// each suite self-checks before the timed loop
if (await verify(vp) !== true)
  throw new Error('self-check failed')
\end{lstlisting}

\balance

\end{document}